\documentclass[prb,twocolumn,showpacs,preprintnumbers,amsmath,amssymb,superscriptaddress,footinbib]{revtex4-1}

\pdfoutput=1

\usepackage[export]{adjustbox}
\usepackage{amsfonts,amsmath,times,mathrsfs,graphicx,dcolumn,bm,wasysym,xcolor,
epsfig,amssymb,lipsum,color,epstopdf,bbold}
\usepackage[makeroom]{cancel}
\usepackage[english]{babel}
\usepackage[T1]{fontenc}
\usepackage[colorlinks,bookmarks=true,citecolor=red,linkcolor=blue,urlcolor=blue]{hyperref}
\usepackage{hyperref}

\usepackage[tight, FIGTOPCAP, hang, raggedright, nooneline]{subfigure}
\usepackage{xcolor}
\definecolor{Green}{RGB}{80,182,0} 
\begin{document}
\setlength{\parskip}{0pt}

\title{Emergent statistical bubble localization in a $\mathbb{Z}_2$ lattice gauge theory}

\author{H. Yarloo}
\affiliation{Department of Physics, Sharif University of Technology, P.O.Box 
11155-9161, Iran}
\author{M. Mohseni-Rajaee}
\affiliation{Department of Physics, Sharif University of Technology, P.O.Box 
11155-9161, Iran}
\author{A. Langari}
\email{langari@sharif.edu}
\affiliation{Department of Physics, Sharif University of Technology, P.O.Box 
11155-9161, Iran}

\begin{abstract}
We introduce a clean cluster spin chain coupled to fully interacting spinless fermions, forming an unconstrained $\mathbb{Z}_2$ lattice gauge theory (LGT) which possesses dynamical proximity effect controlled by the entanglement structure of the initial state. 
We expand the machinery of interaction-driven localization to the realm of LGTs such that for any starting product state, the matter fields exhibits emergent statistical bubble localization, which is driven solely by the cluster interaction, having no topologically trivial non-interacting counterpart, and thus is of a pure dynamical many-body effect. 
In this vein, our proposed setting provides possibly the minimal model dropping all the conventional assumptions regarding the existence of many-body localization. Through projective measurement of local constituting species, we also identify the coexistence of the disentangled nonergodic matter and thermalized gauge degrees of freedom which  
stands completely beyond the standard established phenomenology of quantum disentangled liquids. 
As a by product of self-localization of the proximate fermions, the spin subsystem hosts the long-lived topological edge zero modes, which are dynamically decoupled from the thermalized background $\mathbb{Z}_2$ charges of the bulk, and hence remains cold at arbitrary high-energy density. This provides a convenient platform for strong protection of the quantum bits of information which are embedded at the edges of completely ergodic sub-system; 
the phenomenon that in the absence of such proximity-induced self-localization could, at best, come about with a pre-thermal manner in translational invariant systems.
Finally, by breaking local $\mathbb{Z}_2$ symmetry of the model, we argue that such admixture of particles no longer remains disentangled and the ergodic gauge degrees of freedom act as a "small bath" coupled to the localized components.

\end{abstract}


\maketitle

\section{Introduction}\label{sec:introduction}
It was taken for granted that the presence of interaction would bring ergodicity back to a single-particle localized system; until one decade ago, when it was argued by Basko, Aleiner, and Altshuler (BAA)~\cite{Basko:2006}, 
that Anderson localization (AL)~\cite{Anderson:1958} is stable under weak short range interactions, a phenomenon known as many-body localization (MBL)~\cite{Gornyi:2005,Basko:2006,Oganesyan:2007,Pal:2010}. Existence of MBL demurs against the universal applicability of conventional statistical mechanics in a closed quantum system by violating the "eigenstate thermalization hypothesis"~\cite{Deutsch:1991,Srednicki:1994,Rigol:2008}. Extensive theoretical, numerical, and experimental studies have examined various aspects of such generic alternative to thermalization, highlighted by 
tenacity of retrievable information after arbitrary long time~\cite{Serbyn:2014_1,Serbyn:2014_2}, unbounded logarithmic growth of entanglement entropy~\cite{Znidaric:2008,Bardarson:2012} as well as its area law finite-size scaling even in highly excited states~\cite{Bauer:2013,Luitz:2015}. The latter together with the absence of thermalization are the most remarkable features of MBL, making it a viable platform for protection of quantum order all the way to infinite temperature~\cite{Huse:2013,Chandran:2014,Bahri:2015,Potter:2015,Yao:2015}.

Following the original work of BAA, majority of researches evince the existence of MBL in a closed quantum system through the presence of two key ingredients: \textit{quenched disorder} in the Hamiltonian and localization of \textit{all} single-particle states. That being the case, a generic MBL Hamiltonian takes a diagonalized form in the "l-bit" basis constructed by an extensive set of local integrals of motion (LIOM), heralding the emergent integrability of the system~\cite{Serbyn:2013_2,Huse:2014,Chandran:2015,Rademaker:2016,Imbrie:2016}. In the deep MBL regime, these IOMs resemble their non-interacting counterparts and could be viewed as locally dressed density operators of Anderson orbitals~\cite{Rademaker:2016,Imbrie:2016}. In the conventional wisdom, every MBL eigenstate can be continuously deformed via a finite depth quantum circuit to the one corresponded to an Anderson insulator~\cite{Huse:2014,Bauer:2013}. This phenomenological picture tends to make one think of Fermi-liquid like adiabatic connection which holds in the entire spectrum of such conventional MBL (cMBL) phase~\cite{Bera:2015,Bera:2017}. However, the basic requirements of MBL are going to be modified.

Recently, the presumed necessity of the outlined ingredients for the presence of localization is a subject of ongoing debate~\cite{Li:2015,Li:2016,Bar:2016,Li:2017,Schiulaz:2015,Papic:2015,Yao:2016,Yarloo:2018,Smith:2017,Smith:2017-b,Brenes:2018,Schulz:2018}. A fresh perspective proposed an emergent dynamical glassiness, initiated in an inhomogeneous state, which is realizable in translational invariant models consisting of mass-imbalanced (quasi-) particles~\cite{Schiulaz:2015,Papic:2015,Yao:2016,Yarloo:2018}. In spite of their eventual thermalization, in the intermediate time scales they look like true MBL, hence the name "quasi-MBL"~\cite{Yao:2016}. 
Such dynamical localization yet with topological non-trivial character is also reported in a self-correcting memory, with the mutual braiding statistics as the origin of the emergent disorder~\cite{Yarloo:2018}.
A closely related concept also concerns the coexistence of disentangled localized and ergodic degrees of freedom in multi-component systems which brings about the idea of quantum disentangled liquid (QDL)~\cite{Grover:2014,Garrison:2016}; a thermal state of matter whose hidden locality can be uncovered through a projective local measurement of different constituting species. The first microscopic Hamiltonian exhibiting such partial breakdown of thermalization has been identified in the Hubbard-like model whose low energy spectrum hosts the admixture of spin and charge excitations~\cite{Garrison:2016}.

Moreover, a number of recent studies proposed lattice gauge theories (LGTs) exhibiting dynamical localization of the matter fields through the gauge super-selection rules~\cite{Smith:2017,Smith:2017-b,Brenes:2018}. The idea behind such disorder-free localization can be traced back to an early work of Ref.~\onlinecite{Paredes:2005} on quantum disorder simulation protocol. There, the dynamics of a \textit{binary disordered} system could be captured by introducing ancillary degrees of freedom into a \textit{clean} model, whereby the influence of disorder is effectively simulated via quantum superposition. 
Unlike the quasi-MBL, localization of this type is not transient and requires no configurational disorder within the initial states. In these LGTs, the dynamics of the localized sub-system displays some facets of cMBL in the way that the interaction between one type of particles serves as a perturbation which tends to drive the whole system towards thermalization. Moreover in the \textit{non-interacting} framework, as proposed by Ref.~\onlinecite{Smith:2017-b}, such LGTs (with factorized spectrum) can behave as QDL, in the sense that they host localized matter in coexistence with the ergodic gauge fields. 

In the whole outlined mechanisms, the emergent cMBL (and also quasi-MBL) dynamics can still be adiabatically connected to that of an effective single-particle localized system, e.g., Anderson~\cite{Anderson:1958} or Wannier-Stark insulator~\cite{Wannier:1962}. An engaging question which arises from these studies is whether generic non-ergodic dynamics can be found in a minimal model dropping all the mentioned essential ingredients needed for the existence of the cMBL and how much robust the resulting dynamical glassiness is to generic perturbations? 

Here, we address these questions by constructing a clean $\mathbb{Z}_2$ LGT at finite fermion density through coupling a cluster spin chain to interacting spinless fermions. 
Given the symmetry protected topological (SPT) character of the cluster chain, we argue that a quantum quench from any arbitrary product state leads to the interaction-driven self-localization of the gauge invariant fermions, in a model whose non-interacting dynamics is completely extended. Without adiabatic connection to AL dynamics, this new class of localization, in addition to being disorder-free, is of pure many-body effect.
Indeed, the effective Hamiltonian dynamics of the matter fields in the present model, instead of cMBL, is described by a more recently proposed mechanism known as statistical bubble localization (SBL)~\cite{Li:2017}. Originally, the idea of the SBL was introduced as a non-perturbative approach toward disorder-induced localization, which despite standing completely beyond the LIOM picture, shares many similarities with cMBL.
Our results confirm the fingerprints of the SBL in the matter's dynamics, where the tenacity of emergent localization relies directly on the global fermionic pattern of the pre-quench states. Besides, by tuning the interaction strength we unveil a qualitative change in dynamics, from emergent SBL in strongly interacting limit, to an anomalous extended slow behavior~\cite{Agarwal:2015,Herrera:2015,Lev:2015,Luitz:2016,Znidaric:2016,Khemani:2017,Doggen:2018} at the ergodic side of dynamical transition point, impressively, in the absence of both definable single-particle localization length and spatially atypical regions in the initial state.

Moreover, we argue that the non-ergodic matter fields in this two-component LGT are embedded throughout the fully thermalizing background $\mathbb{Z}_2$ charges. However, the triviality of the non-interacting limit aims at making such coexistence to exceed the scope of the QDL phenomenology. 
Instead, the mentioned proximity of localized and ergodic components is reminiscence of the situation giving rise to the cMBL proximity effect~\cite{Nandkishore:2015,Hyatt:2017,Marino:2018}; the phenomena in which localization can be induced in the ergodic bath through coupling to a cMBL system with comparable number of degrees of freedom. Here we present a new class of proximity effect, in which the localization of fermions dynamically leaks to the topological edge zero modes of the cluster chain, preventing them from rapidly ending up in a trivial state, while at the same time, the rest of spin modes are utterly thermalized. In this regard, initially stored quantum information can be protected at the edges of the clean and completely ergodic subsystem.

Finally, in the presence of generic gauge spoiling perturbations 
the previously disentangled ergodic and localized subsystems become tangled, through which the ergodic nature of the spin subsystem can be revealed by its bath-like effect on the fermions. In the light of this, the dynamics demonstrates lingering signatures of localization with a stretched slow relaxation of time-correlators accompanied by an exponentially slow heating of the whole system, leading up to inevitable thermalization at the late times.

The paper is structured as follows: In Sec.~\ref{sec:model} we introduce the model and describe its dynamical equivalence to SBL, which forms the basis for the reminder of the paper. Sec.~\ref{sec:interaction-driven} describes in detail our numerical approaches and the resulting data corresponding to the dynamics of fermions, spins and entanglement entropy. Sec.~\ref{sec:bath-like} focuses on the robustness of dynamical SBL under generic gauge spoiling perturbations. We conclude the paper by briefly summarizing our main results with discussions in Sec.~\ref{sec:discuss}.

\section{$\mathbb{Z}_2$ lattice gauge theory for the cluster spin chain}\label{sec:model}

\begin{figure}[t!]
\centering
\includegraphics[width=0.95\linewidth]{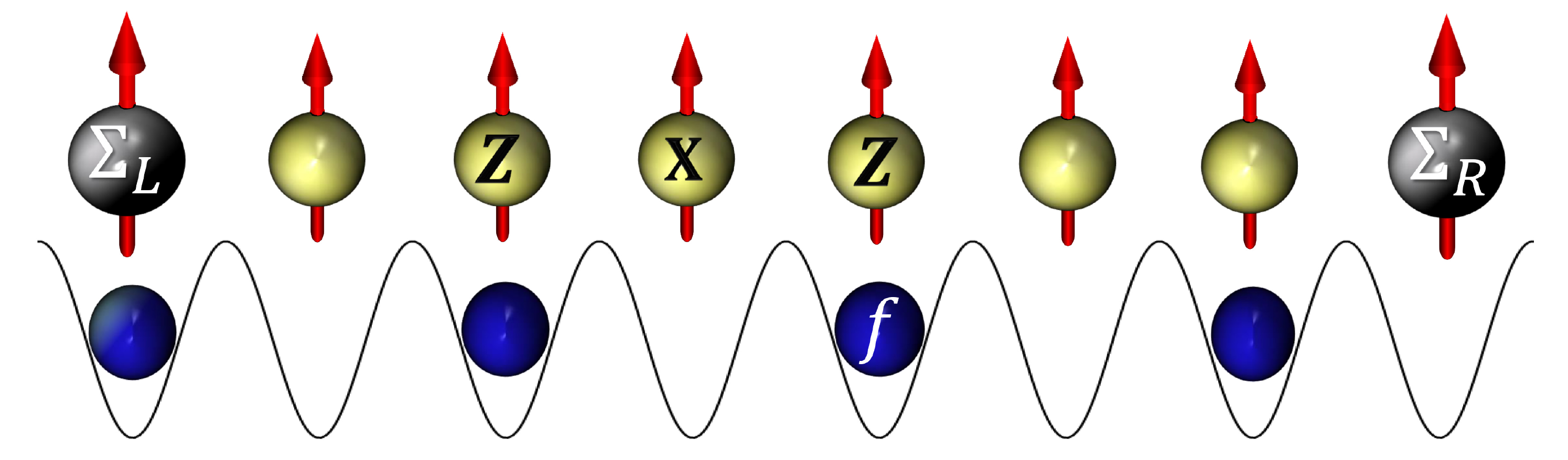}
\caption{Schematic representation of the cluster spin chain (upper spheres) coupled to spinless fermions (lower spheres), given by the translationally invariant Hamiltonian~(\ref{eq:H_cf}). The black spheres, at the ends of spin chain, indicate the topological edge qubits defined in Eq.~(\ref{eq:soft_edges}). 
}
\label{fig:model} 
\end{figure}

Our starting point is the clean cluster Hamiltonian,
\begin{align}
    H_{Cluster}=\lambda \sum_i \mathcal{W}_i + h \sum_i X_{i} + V \sum_{\langle ij\rangle} X_{i}X_{j}.
       \label{eq:H_cluster}
\end{align}
The first term reads as the sum of cluster stabilizer terms $\mathcal{W}_i= Z_{i-1}X_{i}Z_{i+1}$, where $Z_i$ ($X_i$) is the $z$ ($x$) component of the spin-$1/2$ positioned on site $i$ of a chain with length $N$. 
The cluster spin chain is shown in the upper part of Fig.~\ref{fig:model}, where
each spin is shown by a sphere containing an upward arrow.

The eigenstates of the cluster term are defined by $\mathcal{W}_i |\{ w_i \pm \}\rangle = w_i |\{ w_i \pm \}\rangle$ with $w_i = \pm 1$, which have non-trivial multipartite entanglement structure. 
The cluster eigenstates have the following form
\begin{align}\label{eq:cluster_state}
|\{ w_i \pm \}\rangle = e^{i\frac{\pi}{4}(Z_1 +Z_N)} e^ {i\frac{\pi}{4} \sum_{i=2}^{N-1} Z_i Z_{i+1}}|x \pm \rangle^{\otimes N},
\end{align}
where $|x \pm \rangle$ is the eigenstate of $X$, i.e., $X|x \pm \rangle = \pm|x \pm \rangle$. 
They are the archetype of $\mathbb{Z}_2\times\mathbb{Z}_2$ SPT order, having an exact matrix product state (MPS) expression which cannot be smoothly connected to a product state without explicit or spontaneous breaking of the Ising-like symmetry $P_{e/o}=\prod_{i \, \in\, e/o} X_i$, defined on even and odd lattice sites, separately. 

The first two terms of the Hamiltonian~(\ref{eq:H_cluster}) can be mapped to a free Majorana fermion model by a non-local Jordan-Wigner transformation, and the last term makes this model generic. The topological nature of the model is manifested through the appearance of a pair of Majorana zero modes (MZMs) at each end of the chain. In this respect, one can introduce the edge operators,
\begin{align}\label	{eq:soft_edges}
    \Sigma^z_L= Z_{1} ,\quad \Sigma^x_L= X_{1} Z_{2}  ,\quad \Sigma^y_L= Y_{1}Z_{2},
\end{align}
at the left end (and at the right as well, see Fig.~\ref{fig:model}), which commute with the $\mathcal{W}_i$'s, are odd under $P_{e/o}$ and obey the Pauli algebra. In the non-interacting limit, these edge zero modes approximately commute with the Hamiltonian in finite size and are decoupled from bulk degrees of freedom.
However, in the generic case, any quantum information encoded in these {\textit{soft}} edge modes is rapidly lost due to the hybridization with the thermal bulk excitations at finite temperature. It has been shown that disordering the Hamiltonian parameters ($\lambda_i, h_i, V_i$) in its topological phase, i.e. $\bar{\lambda}_{i} \gg \bar{h}_i$, leads to localization of the bulk modes, which prevents the absorption and emission of them by the zero modes~\cite{Bahri:2015}. By doing so, the initially stored boundary information even at high energy densities (i.e. close to the middle of the spectrum) does not spread trivially through the localized bulk, thereby the coherency of the edge modes would be sustained.

Here, instead of introducing external disorder into the system, we propose to couple the Hamiltonian~(\ref{eq:H_cluster}) to interacting spinless fermions $f_{i}$,
\begin{align}    \label{eq:H_cf}
    H(\lambda,V,h)&=
    \lambda \sum_i\, n_{i-1}\,\mathcal{W}_i\,n_{i+1}\\
    &+ V \sum_{\langle ij\rangle} X_{i}X_{j} \,f^{\dagger}_{i} f_{j} + h \sum_i X_{i}, \nonumber
\end{align}
where $n_{i} \equiv n_{i}^f= f^{\dagger}_{i} f_{i} $ represents the local 
fermionic occupation operator. A schematic picture is shown in Fig.~\ref{fig:model}, where
the (upper) spins are coupled to the (lower) fermions in the lattice, through Eq.~(\ref{eq:H_cf}).
Such settings are also widely considered in other contexts including slave-particle description of the Hubbard model~\cite{Ruegg:2010,Zitko:2015}, orthogonal metals~\cite{Nandkishore:2012,Prosko:2017} and non-Fermi glasses~\cite{Parameswaran:2017}.

In order to reveal the gauge symmetry and the resulting dynamical localization of the fermion-cluster model~(\ref{eq:H_cf}), we implement the following non-local dual mapping:
\begin{align}\label{eq:dual_mu}
 \left\{
 \begin{array}{ll}
    \hat{\mu}^z_{2i}&=Z_{2i}\prod_{k\leqslant i} X_{2k-1} ,\\ \\
    \hat{\mu}^z_{2i-1}&=Z_{2i-1}\prod_{k\geqslant i} X_{2k},
 \end{array}
 \quad, \,\,\,\, \hat{\mu}^x_{i}=X_i.
\right.
 \end{align}
In the $\hat{\mu}_{i}$ representation, we can define the new $c_i$ fermions that commute with the $\hat{g}_i$ operators which in turn obey the spin-$1/2$ Pauli algebra:
\begin{align} \label{eq:gc}
     \hat{g}^z_i = \hat{\mu}^z_{i} (-1)^{n_i}, \quad \hat{g}^x_i = \hat{\mu}^x_{i}=X_i, \quad  c_i = \hat{\mu}^x_{i} f_i,
\end{align}
and thus the Hamiltonian~(\ref{eq:H_cf}) takes the form of {\it{fully interacting}} fermions coupled to two decoupled transverse field Ising models in the language of $\hat{g}$-spins,
\begin{align}\label{eq:H_gc}
    H(\lambda,V,h)=
    \lambda \sum_{\ll ij \gg}  n_i\, \hat{g}^z_{i}\, \hat{g}^z_{j}\, n_j
    + V \sum_{\langle ij \rangle} c^{\dagger}_{i} c_{j} + h \sum_i \hat{g}^x_{i},
\end{align}
where $n_i \equiv n_i^c=n_i^f$. The Hamiltonian~(\ref{eq:H_gc}) has a global $U(1)$ particle number conservation for $c$-fermions, 	
in addition to the global $\mathbb{Z}_2\times\mathbb{Z}_2$ (Ising) symmetry which is now translated to $P_{e/o} =\bar{P}_{e/o}\equiv \prod_{i \, \in\, e/o} \hat{g}^x_i$. 

\subsection{Static gauge field: $h=0$ }\label{subsec:h0}
In the absence of an external field, the $\hat{g}^z_{i}$ operators behave as classical variables $\hat{g}^z_{i}\equiv {g}_i$,  and thus construct an extensive set of IOMs, $[\hat{g}_i,\hat{H}]=0$.  Despite the extensive number of such conserved quantities, this model is not generically integrable since total number of degrees of freedom, after including the $N$ number of conserved ${g}_i$s, still remain extensive. These conserved charges are associated to the local $\mathbb{Z}_2$ symmetry implemented by the unitary operator,
\begin{align} \label{eq:gauge-symmetry}
\hat{G}(\{\eta_i\})=\prod_i {{g}^z_i}^{(1-\eta_i)/2},
\end{align}
where $\eta_i=\pm1$. The action of $\hat{G}(\{\eta_i\})$ in terms of the species involved in Eq.~\ref{eq:H_cf}, is given by the transformations
${X}_i\rightarrow \eta_i {X}_i$ and  ${f}_i\rightarrow \eta_i {f}_i$, conveying that the ${c}$-fermions are the {\textit{gauge-invariant}} excitations of the model. 
One can introduce the background $\mathbb{Z}_2$ field ${\mathcal{U}}_{i} \equiv \hat{g}^z_{i-1} \hat{g}^z_{i+1}$, whereby the Hilbert space of the Hamiltonian
\begin{align}
    H_\mathcal{M}^{\{g_i\}}\equiv H_\mathcal{M}^{\{u_i\}}=\lambda \sum_{i}  n_{i-1} \, \mathcal{U}_{i}\, n_{i+1} + V \sum_{\langle ij\rangle} c^{\dagger}_{i} c_{j},
    \label{eq:H_M}
\end{align}
spans over all background  
charge sectors, inferring an {\it{unconstrained}} lattice gauge theory.
This is in opposition to the constrained gauge-matter theory, 
which only relies on the gauge-invariant physical states of the extended Hilbert space from the restriction enforced through the Gauss' law.
In contrast to the gauge symmetries, here the local $\mathbb{Z}_2$ invariance~(\ref{eq:gauge-symmetry}) is an actual symmetry of $H_\mathcal{M}^{\{u_i\}}$.

The gauge invariant Hamiltonian~(\ref{eq:H_M}) has an additional $U(1)$ symmetry corresponding to the conserved total number of $\mathbb{Z}_2$ charges. This can be revealed by defining a new fermionic field~$\chi$, equivalent to the $\mathbb{Z}_2$ charge imposed by Gauss' law, in terms of which ${\mathcal{U}}_{i} \equiv 1-2n^{\chi}_{i}$ with $n^{\chi}_{i}\equiv\chi^{\dagger}_{i}\chi_{i}=\{0,1\}$. In this respect, $N^{\chi}_{e/o}=\sum_{i\in e/o}n^{\chi}_{i}$ separately counts the number of kink excitations in the two decoupled Ising chains (formed by $\hat{g}^{z}_{i}$) for a given gauge configuration and
$P_{e/o}$ renders as the parity of $\chi$-fermions, separately in even and odd sites.

Besides, since the gauge fields are \textit{static}, all eigenstates of the Hamiltonian~(\ref{eq:H_M}) take the factorized structure,
\begin{align}\label{eq:full_Hilbert}
|\Psi\rangle = | \mathcal{G}\rangle \otimes  | \mathcal{M_{\mathcal{G}}} \rangle,
\end{align}
where the gauge field $|\mathcal{G}\rangle$ is specified by ${\mathcal{U}}_{i}| \mathcal{G}\rangle=u_{i}| \mathcal{G}\rangle$ and the eigenstate of the gauge invariant (fermionic) matter excitations corresponding to a given set of $\{u_{i}\}$ is denoted by $|\mathcal{M_{\mathcal{G}}} \rangle$. Remarkably, the background field enjoys some principal properties. First, despite the fact that the $\hat{g}_i$ operators are of the string type, ${\mathcal{U}}_{i}$ possesses a {\it{local}} structure, which is apparent in terms of the original fermion and spin operators,
\begin{align}\label{eq:U_j}
{\mathcal{U}}_{i}=\mathcal{W}_i\,(-1)^{n_{i-1}+n_{i+1}}.
\end{align}
Second, since $[\hat{\mathcal{U}}_{i},\hat{P}_{e/o}]=0$, all $2^{N-2}$ arrangements of $\{u_{i}\}$ are uniquely identified through only $N-2$ independent $\{w_i\}$, which according to the SPT character of the cluster Hamiltonian, points to a one-to-four correspondence to $\{g_i\}$ configurations. This topological character of $H_\mathcal{M}^{\{u_i\}}$, for any nonzero value of $\lambda$, is attributed to the appearance of new ``edge strong zero modes''~\cite{Kemp:2017,Fendley:2016,Else:2017} at the left end (and similarly at the right one, see Appendix~\ref{App:MZM}), 
\begin{align}\label{eq:strogn_modes}
    \bar{\Sigma}^z_L&= (-1)^{n_1} \Sigma^z_L , \nonumber \\ 
    \bar{\Sigma}^x_L&= (-1)^{n_2} \Sigma^x_L ,\\ 
    \bar{\Sigma}^y_L&= (-1)^{n_1+n_2} \Sigma^y_L, \nonumber
\end{align}
which, strikingly, are local objects in terms of the occupation number of the original fermion and spin edge modes defined in Eq.~\ref{eq:soft_edges}. They obey Pauli algebra and behave the same as the original spins under the $\mathbb{Z}_2$ gauge transformation~(\ref{eq:gauge-symmetry}). Specifically, while the exact commutation relation, 
\begin{align}\label{eq:comm_strogn_modes}
[H(\lambda,V,0), \bar{\Sigma}]=0,
\end{align}
holds, $\bar{\Sigma}_{L/R}$ operators are odd under $\bar{P}_{e/o}$ and thus do not appear explicitly in $H_\mathcal{M}^{\{u_i\}}$. Hence, they are indeed physically addressable degrees of freedom which are effectively decoupled from all the rest and emerge in the composite system owning to the proximity of the cluster chain to the fermion spices. Such modes connect eigenfunctions from different topological sectors with exactly the same energy and the term ``strong'' implies the validity of this mapping for all eigenstates of the Hamiltonian~\cite{Kemp:2017,Fendley:2016,Else:2017}. Thereby, the gauge sector of the full Hilbert space in Eq.~\ref{eq:full_Hilbert} can be further broken up into tensor products of those topologically non-trivial edge qubits (of dimension $2^{2}$) and the gauge invariant background fields of the bulk.
In the Majorana formulation, $\bar{\Sigma}_{L/R}$ can be realized as two pairs of localized strong MZM with the non-Abelian braiding for adiabatic exchange of them~\cite{Ivanov:2001,Clarke:2011}.

Finally, in the presence of gauge spoiling terms, i.e., $h\neq0$, the operators $\hat{g}_i$s become typical spin-$1/2$ Pauli operators defined in Eq.~\ref{eq:gc}. In this way, giving dynamics to the static charges breaks the mentioned $U(1)$ symmetry, and thus the Hamiltonian~\ref{eq:H_cf} would no-longer remain block-diagonal. In Sec.~\ref{sec:bath-like}, we shall thoroughly investigate the effects of such perturbations on the out-of-equilibrium dynamics of the model.

\subsection{Dynamical localization controlled by the entanglement structure of initial states}\label{subsec:DLCBTESOIS}

We now look over the the standard quantum quench protocol by considering initial states of the form,
\begin{align}\label{eq:psi0}
 |\Psi_0\rangle=|\psi_0^f\rangle\otimes|\psi_0^\sigma\rangle,
\end{align}
as the source of quasi-particle excitations, where $|\psi_0^f\rangle$ ($|\psi_0^\sigma\rangle$) is the initial state of the fermion (spin) degrees of freedom. Here, the fermions are prepared in local occupation numbers described by the Slater determinant products, namely, a charge density wave (CDW) $|\cdots1010\cdots\rangle$ and a domain-wall (DW) $|\cdots1100\cdots\rangle$. We establish that dynamics of the fermionic subsystem is dictated by the {\it{entanglement structure}} of the initial spin states. To this end, we refer to the two inclusive families of pure states in 1D, which they fall into: un-entangled arbitrary product states and short-range entangled cluster states given by Eq.~\ref{eq:cluster_state}. Following Eq.~\ref{eq:U_j}, any arbitrary spin state of the former family, $|\psi_0^\sigma\rangle=\otimes_{i=1}^{N}|\sigma_i\rangle$, under the action of $\mathcal{U}$ transforms to an equal superposition of $2^{N-2}$ gauge field configurations $\{u_i\}$ (equivalent to $2^N$ charge configurations $\{g_i\}$), which dynamically reads as:
\begin{align}\label{eq:quench}
  |\Psi(t)\rangle=e^{-iHt}|\Psi_0\rangle=\frac{1}{\sqrt \mathcal{N}}\sum_{\{u_i\}} e^{-i H_\mathcal{M}^{\{u_i\}} t}|\mathcal{M_{\mathcal{G}}}\rangle,
\end{align}
with $\mathcal{N}$ specifying $2^{N-2}$ distinct super-selection sectors. Since $n_i^f=n_i^c$, both the fermion and matter degrees of freedom have identical occupation representation.

\begin{figure}[t!]
\centering
\includegraphics[width=0.95\linewidth]{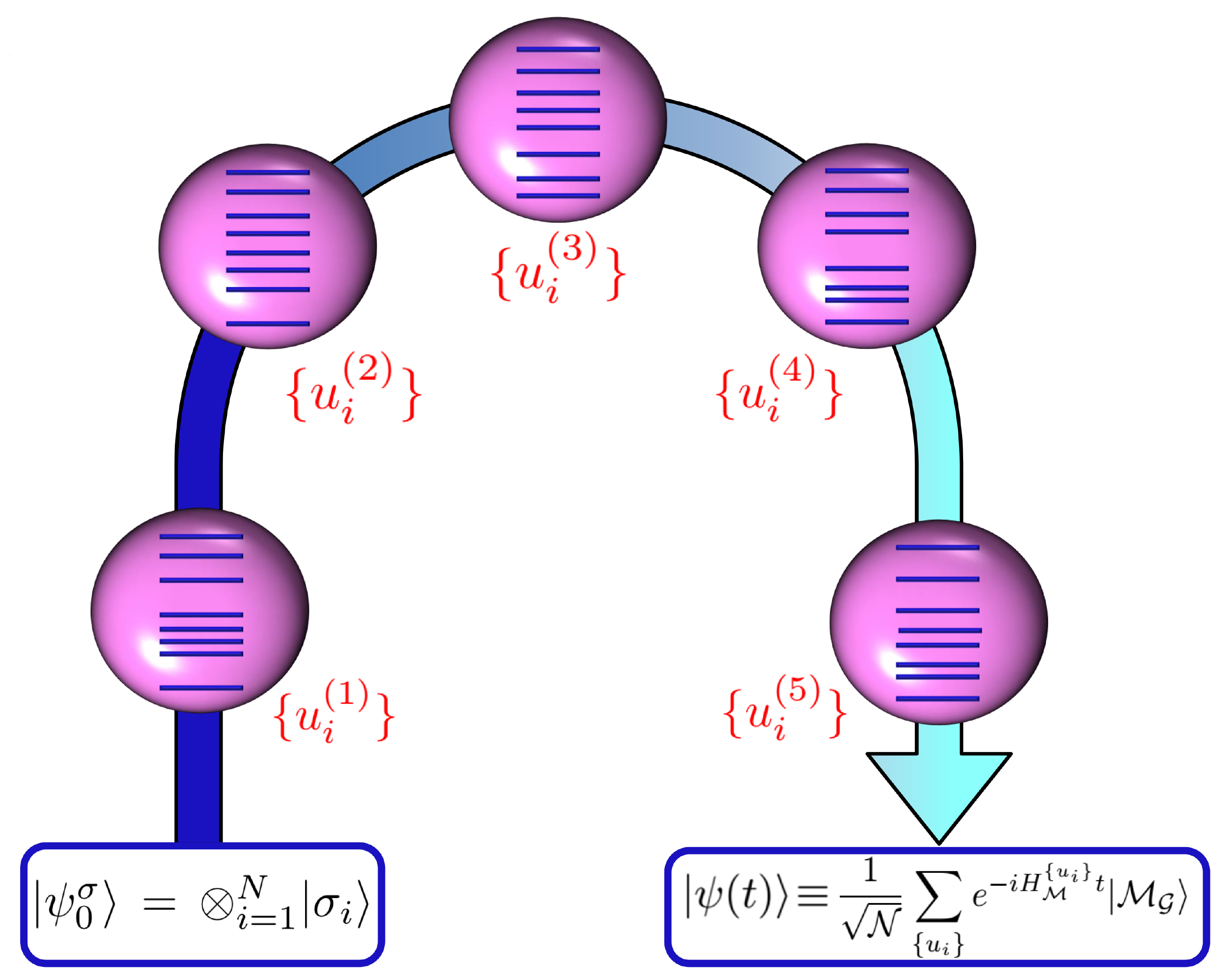}
\caption{Illustrative representation of system dynamics starting from an initial state, which is a product state in spin sector. Dynamically generated different configurations of spins act as disorder on matter sector. }
\label{fig:configurations} 
\end{figure}

In this way, by starting from an arbitrary initial spin product state, the matter dynamically faces all the super-selection sectors and will evolve in each of them according to unalike Hamiltonians $H^{\{u_i\}}_{\mathcal{M}}$, as shown in Fig.~\ref{fig:configurations}.
Thus, after integrating out the gauge fields, dynamical traits of the matters can be simply recast under,
\begin{align} \label{eq:disordered_H}
H^{\{\lambda_i\}}_{\mathcal{M}}=  V \sum_{\langle ij \rangle} c^{\dagger}_{i} c_{j} +\sum_{\ll ij \gg} \lambda_i\,n_i n_j,
\end{align}
where the local random sign of the distinctive static charge configurations provides a {\it{random-interaction}} pattern with the strength of 
$\lambda$. In this regard, initializing system in any arbitrary spin product structure leads to a dynamical glassiness without the influence of an \textit{emergent on-site disorder potential}. 
The ensuing {\it{dynamical}} localization under $H^{\{\lambda_i\}}_{\mathcal{M}}$ is driven solely by {\it{interaction}} and thus is of a pure many-body effect. Indeed, the matter degrees of freedom in the single particle limit $\lambda=0$ (equivalent to topologically trivial regime of the spin subsystem) are totally extended. Therefore, in stark contrast to the previous studies on disorder-free localization~\cite{Smith:2017,Smith:2017-b,Michailidis:2017,Schulz:2018} and quasi-MBL in transnational invariant systems~\cite{Schiulaz:2015,Papic:2015,Yao:2016,Yarloo:2018}, here, in the limit of vanishing interaction, the matter dynamics is not adiabatically connected to an effective single-particle localized model. Hence, our proposed setting for interaction-driven self-localization is possibly the minimal (gauge) model dropping all prior expectation regarding cMBL.

In contrast to the case of initial spin product state, dynamics starting from the short-range entangled cluster state, defined in Eq.~\ref{eq:cluster_state}, maps to that of fermions in a fixed (uniform or non-uniform) configuration of background $\mathbb{Z}_2$ charges. From the above perspective, the initial entanglement structure of the spin subsystem, which in essence is non-local, can be recollected through the non-ergodic dynamics manifested in some local fermionic observables. These should be compared with the earlier studies on disorder-free localization in LGTs~\cite{Smith:2017,Smith:2017-b,Brenes:2018} in which the initial polarization of each spin, as a local property, determines matter's dynamics.

\subsection{Connection to SBL}

It has been recently shown that in the presence of \textit{external disorder} random-interaction Hamiltonians of the form Eq.~\ref{eq:disordered_H} give rise to the so-called SBL~\cite{Li:2017} which, opposed to the on-site disorder cMBL, cannot be phenomenologically described by LIOMs. In the strong randomness limit, the SBL eigenstates form a ``bubble-neck'' structure embracing localized insulating blocks (fermionic clusters with more than one fermion on adjacent sites) along with ergodic thermal bubbles (clusters with isolated fermions). In contrast to the Fermi-liquid like eigenstates of the cMBL with short-range entanglement~\cite{Bauer:2013,Luitz:2015}, bubble-neck eigenstates could be long-range (volume-law) entangled, even though, in the thermodynamic limit they could be area law on average~\cite{Li:2017}.

Due to the dynamical origin of the glassiness in our model, (and also the fate of LIOMs description for the effective Hamiltonian dynamics~\ref{eq:disordered_H}), our proposed model should dynamically reflect the fingerprints of the SBL in the sense that the tenacity of interaction-driven self-localization relies directly on the global pattern of the primary fermionic states, as an alternative to their energy variance. We will track the manifestation of this feature through the partial occupations of quasi-particles, lack of QDL description as well as the stability of the topological spin edge qubits in proximity with the localized bulk fermions.

\section{dynamical localization driven by interaction} \label{sec:interaction-driven}
We now present our extensive numerical results and explore further details regarding the Hamiltonian dynamics $H(\lambda,V,0)$. 
First, since the dynamics of the $c$ (and thus $f$) fermions are invariant with respect to any arbitrary initial spin product states
, from now on we mainly consider tensor products of $z(x)$-polarized spins and fermionic CDW/DW as the initial states, unless otherwise specified. By choosing such translationally invariant initial states, one as well rules out quasi-MBL mechanisms wherein randomness is inherited from inhomogeneous starting states~\cite{Schiulaz:2015,Papic:2015,Yao:2016,Michailidis:2017,Yarloo:2018}. Additionally, we restrict the study to the fermion-cluster chain with open boundary condition; therefore avoiding the {\it{artificial}} excess of quasi-degeneracies as the consequence of picking periodic boundary conditions in systems with binary distribution of randomness\cite{Janarek:2018}. Hereunder, $L=2N$ denotes the number of lattice sites for a chain of $N$ spins coupled to fermions at half-filling.
\subsection{Numerical methods} \label{subsec:Numerical}

For the dynamical simulation of the model at $L<32$, we work under the original model $H(\lambda,V,0)$, defined in Eq.~\ref{eq:H_cf}, by taking advantage of its block-diagonal form in exact diagonalization (ED). For $L\geq32$ we use the effective disordered Hamiltonian~(\ref{eq:H_M}) in which the convergence of data is achieved by averaging over up to $6000$ disorder realizations depending on system size. The provided results in this case are obtained by time-integration based on the massively parallel Chebyshev expansion~\cite{petsc-user-ref,petsc-efficient,Hernandez:2005}, which boasts faster convergence than the Krylov subspace projection based methods. Alternatively, in order to calculate two-point correlation functions in the form of Eq.~\ref{eq:OPDM}, we employed the time-dependent variational principle (TDVP)~\cite{Jackiw:1979,Haegeman:2011,Haegeman:2013,Haegeman:2016} by simultaneous implementation of bosonic and fermionic MPS representations~\cite{Wall:2015,Jaschke:2018} for the original model~(\ref{eq:H_cf}) at $L=32$.

The TDVP method generally prompts non-linear evolution of the wavefunction due to projection into tangent space of MPS manifold with (fixed) finite bond dimension $D$~\cite{Haegeman:2011,Haegeman:2016}. The underlying TDVP trajectories 
effectively generate chaotic classical dynamics~\cite{Haegeman:2011,Leviatan:2017} which crucially obeys the macroscopic symmetries of the original quantum Hamiltonian, regardless of performing the truncation procedure~\cite{Haegeman:2013}. In this respect, the correct hydrodynamic description of macrostates in the {\it{ergodic}} systems is believed to be captured~\cite{Leviatan:2017}, even when one allows only for a low entangled state with a small bond dimension. However, the long-time dynamics extracted in this manner, might be highly misleading as the physical reliability of hydrodynamic observables, in general, could not necessarily be warranted by checking for convergence with such small $D$~\cite{Kloss:2018}. 

Here, we are simulating the exact quantum dynamics for a system containing \textit{at least} one ergodic species. By hiring $U(1)$ symmetry in the fermionic sub-space of MPS manifold we have reached maximum bond dimensions $D_{max}\approx7500$, while setting $\delta t=0.05$ and the absolute error $\varepsilon$ to be smaller than $10^{-7}$. At this highest possible bond dimensions, one expect the $U(1)$ symmetric TDVP to grant the accurate quantum many-body dynamics prior to the Lyapunov time (the time beyond which the classical chaos would be emerged due to considering finite $D$) which itself becomes longer by increasing the bond dimension~\cite{Leviatan:2017}.

\subsection{Dynamics of the fermionic subsystem} \label{subsec:fdynamics}
 \subsubsection{Charge-density wave dynamics}
\begin{figure}[t!]
\centering
\includegraphics[width=0.7\linewidth]{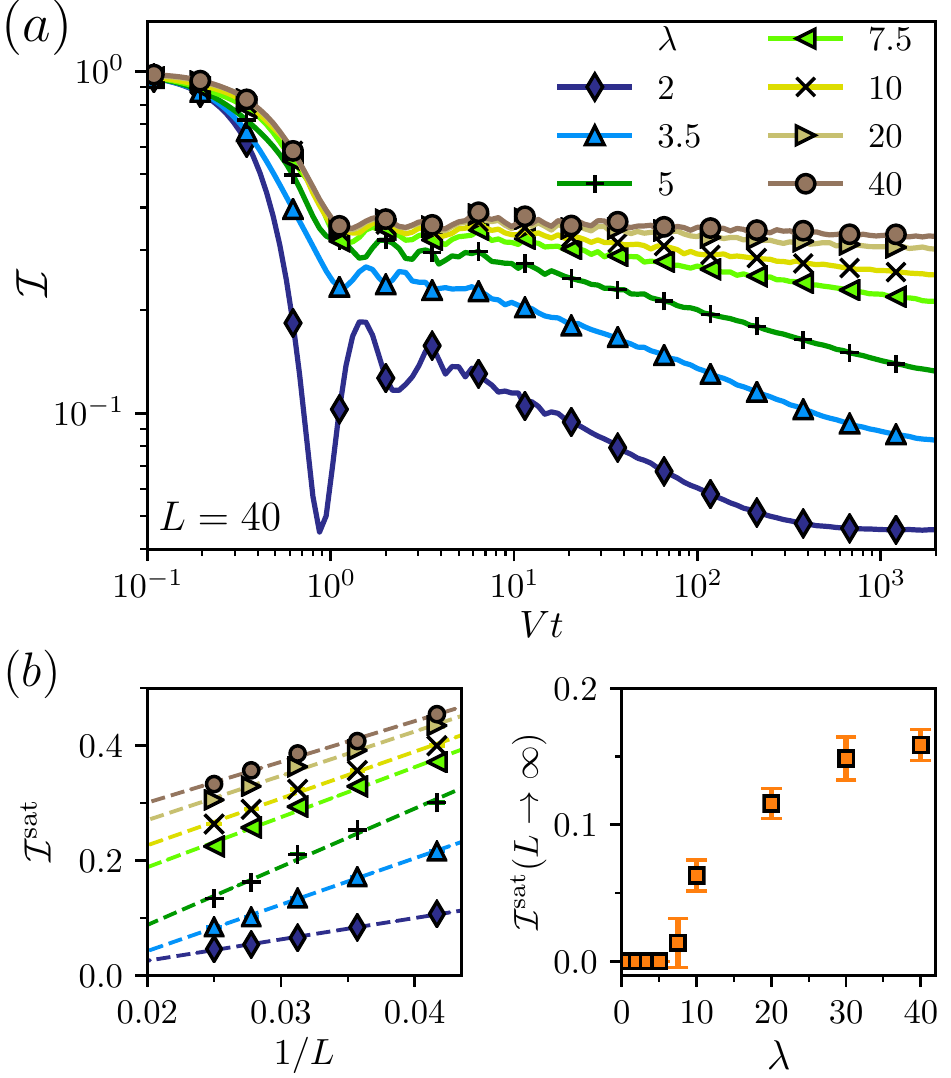}
\caption{(a) Dynamics of the fermionic density imbalance for varying interaction strength $\lambda$ at $L=40$, starting from the initial CDW state. (b) Left panel: Finite-size scaling of the late time value of the imbalance 
for system of sizes $L=24,28,32,36,40$. Right panel: The extrapolated value of the residual imbalance $\mathcal{I}^{sat}(L\rightarrow\infty)$ as a function of interaction.}
\label{fig:inhomo_density} 
\end{figure}

We prepare initially a CDW state and track the dynamics of the $f$-fermions through time evolution of the imbalance $\mathcal{I}=(N^f_e-N^f_o)/(N^f_e+N^f_o)$, where $N^f_e$ ($N^f_o$) denotes the total fermionic occupation at even (odd) sites. As shown in Fig.~\ref{fig:inhomo_density}a, for weak interaction strength, the imbalance experiences an ergodic dynamics and eventually relaxes to its equilibrium zero (finite-size) value. Upon increasing the interaction strength, this relaxation systematically slows down. In the strongly interacting regime the imbalance rapidly approaches a finite stationary value, suggesting the persistence of the microscopic details of the initial state as a result of the emergent SBL phase.
We also pinpoint the SBL transition by considering finite-size scaling of the saturation value of $\mathcal{I}(t)$. The finite offset of its extrapolated value $\mathcal{I}^{sat}_{L \rightarrow \infty}$ is shown to be a convenient dynamical touchstone for the thermal-MBL transition~\cite{Schreiber:2015,Choi:2016,Mierzejewski:2016,Luschen:2017}. In this respect, Fig.~\ref{fig:inhomo_density}b specifies $\lambda_c \approx 7.5$ as the lower bound for SBL transition.

\begin{figure}[t!]
\centering
\includegraphics[width=1\linewidth]{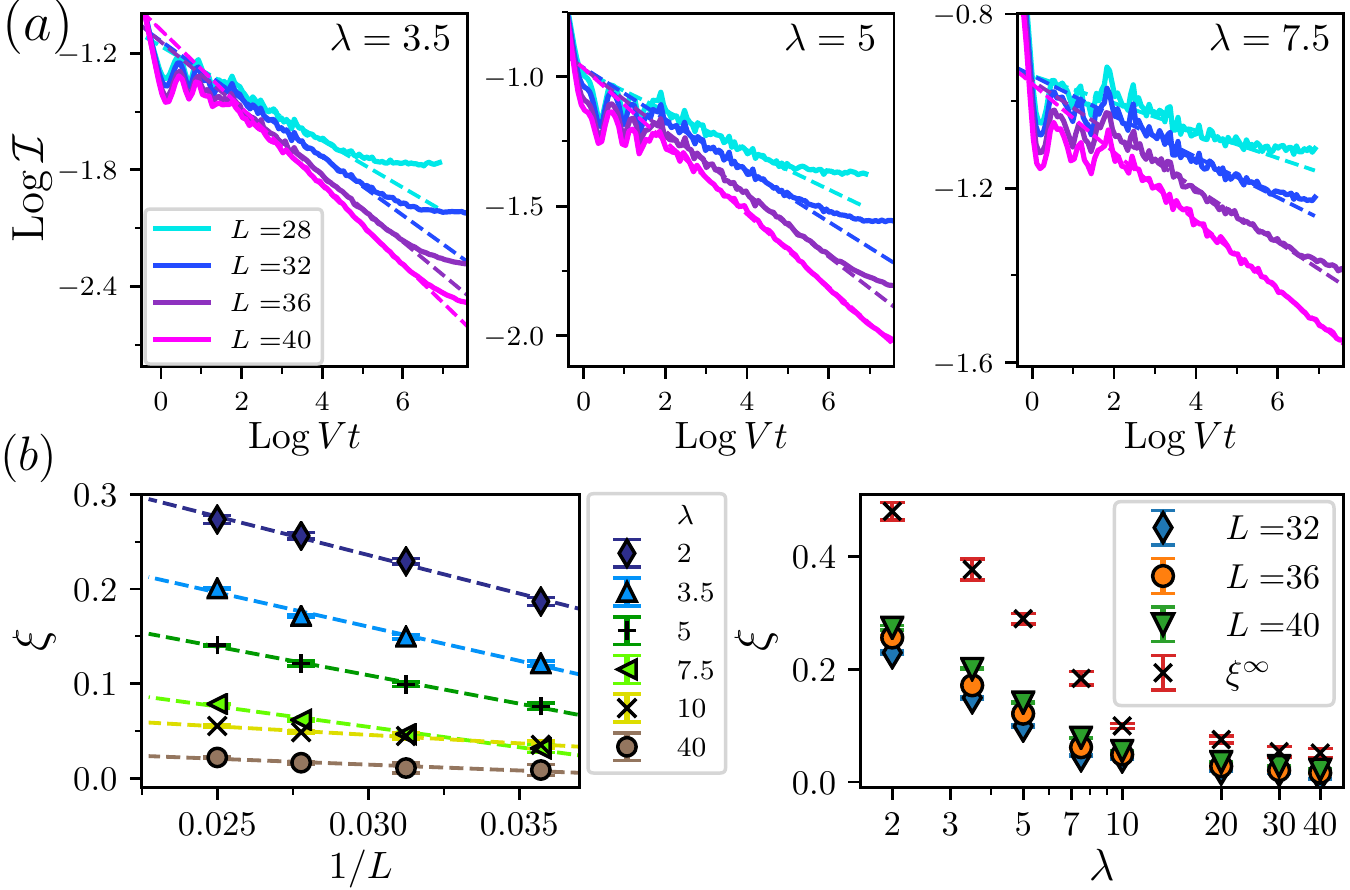}
\caption{(a) Power-law decay of the initial imbalance to its equilibrium value for different interaction strengths in the ergodic regime. The corresponding fitting windows are chosen by excluding the initial sinusoidal damping and the late times where the finite size effects sets in. (b) Left panel: the linear dependence of the decay exponent $\xi$ on $1/L$, extracted for system sizes $L=28,32,36,40$. Right panel: The monotonous decay of $\xi$ by increasing interaction. $\xi^{\infty}$ denotes the extrapolation to $L\rightarrow\infty$.}
\label{fig:Extended_slow} 
\end{figure}

Upon approaching $\lambda_c$ from the ergodic side, for a wide range of interaction parameters, we typify the anomalous extended slow approach to a finite-size long-time value via power-laws $\mathcal{I}(t)\propto t^{-\xi}$ which is akin to those observed in studies on cMBL (with both uncorrolated~\cite{Agarwal:2015,Herrera:2015,Lev:2015,Luitz:2016,Znidaric:2016,Khemani:2017,Doggen:2018} and quasiperiodic~\cite{Bordia:2017,Luschen:2017,Lev:2017,Lee:2017,Khemani:2017} on-site disorder potentials). As demonstrated by the linear behavior in the Log-Log plot, Fig.~\ref{fig:Extended_slow}a, by increasing the system size a wider time window for the power-law decay is observed at intermediate time scales. 
However, the investigation of whether this anomalous slow relaxation of fermionic observables 
comes along with expected sub-linear growth of entanglement entropy~\cite{Luitz:2016,Lev:2017}, is rather tricky. This stems from the dynamical origin of localization and its complexity due to the presence of an ergodic sub-system in our composite model. Nonetheless in Sec.~\ref{sec:Sent_dynamics}, we further confirm the presence of such extended slow regime through sub-ballistic entanglement dynamics of the {\it{matter fields}}.

Additionally, the finite-size scaling of the exponent remains fairly constant in the localized regime. Indeed, by enhancing the interaction strength, the extracted power-law exponent $\xi$ monotonously decreases for different system sizes (see Fig.~\ref{fig:Extended_slow}b), and (approximately) fades near the transition to SBL phase where possibly logarithmic relaxation dynamics is expected~\cite{Mierzejewski:2016}. The stark change in the slope of the extrapolated value of the power-law exponent to $L\rightarrow\infty$
by increasing $\lambda$, clearly distinguishes the thermal and SBL dominions (see the right panel of Fig.~\ref{fig:Extended_slow}b).
This should be noted that near $\lambda_c$, we still observe a power-law behavior (albeit with small exponent) which becomes more clear for large system sizes. This feature suggests that $\lambda_c$ is only the lower bound of the transition point, as observed in the more standard scenarios of cMBL~\cite{Khemani:2017,Khemani:2017-prx,Lee:2017}.
   
\subsubsection{Domain wall dynamics} 
In order to characterize the imprints of primary fermionic patterns on the emergent SBL, we turn to investigate the instantaneous and asymptotic behaviors of one-particle density matrix (OPDM) spectrum~\cite{Bera:2015,Bera:2017,Lezama:2017},
\begin{align}{\label{eq:OPDM}}
    \rho^f_{ij}(t)\equiv \langle f^{\dagger}_i f_j\rangle_{\Psi_0,H}  =\langle\Psi(t)|f^{\dagger}_if_j|\Psi(t)\rangle,
\end{align}
where $|\Psi(t)\rangle$ is defined in Eq.~\ref{eq:quench} and $\text{Tr}\,\rho^{f}(t)=N/2$. We label the spectrum of the OPDM $\{n_\alpha(t) \}$, corresponding to the natural orbitals $\{\phi_{\alpha}\}$, with descending order $n_1\geq n_2 \geq \dots\geq n_N$. Such quantity has been utilized as a formal probe to establish the adiabatic connection between cMBL eigenstates and Anderson orbitals~\cite{Bera:2015,Bera:2017}. In view of this, the occupation spectrum of every cMBL eigenstate finds a non-vanishing gap at the ``emergent Fermi edge''~\cite{Bera:2015,Bera:2017} (i.e. $\alpha=N/2$ at half-filling),
\begin{align}
\Delta n = n_{N/2}-n_{N/2+1},
\end{align}
which evokes the discontinuity of the Fermi-liquid occupancy at zero-temperature~\cite{Bera:2015,Bera:2017}. Yet even deep in the localized phase, a continuous steady-state OPDM spectrum--analogous to the finite temperate Fermi-liquid--can be obtained through a global quench from a local product state~\cite{Lezama:2017}. So, by stipulating \textit{individual cMBL eigenstates} as reference states, one can resolve an emergent temperature specified with the amount of quasiparticle excitations with respect to a certain reference eigenstate~\cite{Lezama:2017}.

Following such intuitions, we evaluate the instantaneous behavior of the OPDM spectrum evolving under the original Hamiltonian $H(\lambda,V,0)$. Here we aim at investigating how the initial size of thermal (or equivalently insulating) blocks within the pre-quench fermionic states provides a dynamical proxy to assess the effective temperature.
Indeed, for the initial states, which have maximum weight on the bubble-neck eigenstates with dominant insulating blocks (e.g., pre-quench DW state containing the largest possible insulating block), one anticipates an atypical slow dynamics along with strong tendency toward self-localization, even at weak interactions (self-disorder). The latter behavior becomes more preferable in the thermodynamic limit, following the exponentially decaying probability distribution of long thermal bubbles~\cite{Li:2017}. However, this is not the case for the preparatory fermionic arrangements with the varying extent of the insulating blocks (e.g., CDW pattern as a perfect thermal bubble).

As we argued above, the fermionic dynamics is invariant under any initial spin product state; like so we let the spin initial state to be $z$-polarized. 
We shall compare the two extreme opposite limits by considering the experimentally relevant DW~\cite{Choi:2016,Hauschild:2016} and CDW pre-quench states~\cite{Schreiber:2015,Luschen:2017} and elaborating upon the resulting differences in dynamics attributed to these states.

\begin{figure}[t!]
\centering
\includegraphics[width=1\linewidth]{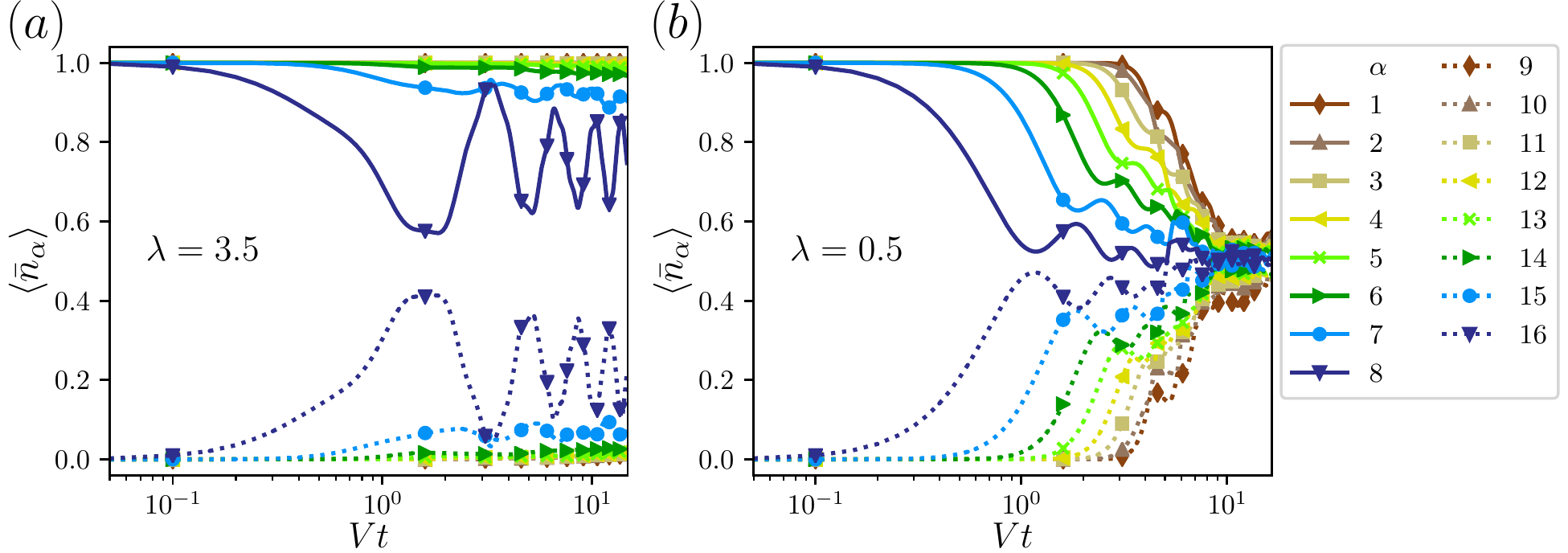}
\caption{The instantaneous behavior of the OPDM spectrum governed by $H(\lambda,V,0)$, following a global quench from DW pattern in the (a) non-ergodic and (b) thermal regimes. The results are provided by $U(1)$ symmetric TDVP with $L=32$ and $D_{max}=7500$. The orbital label is denoted by $\alpha$.} 
\label{fig:3D_TDVP} 
\end{figure}

The results shown in Fig.~\ref{fig:3D_TDVP} correspond to an initial DW configuration as a prefect insulating block. 
At $\lambda = 0.5$ (Fig.~\ref{fig:3D_TDVP}b) the initial discontinuity vanishes with a fast relaxation to an ergodic stationary state. Indeed, the distribution of $n_\alpha$ evolves to a gapless spectrum with all the occupation numbers equal to $n_\alpha \approx 1/2$.
By weakly altering the interaction to $\lambda = 3.5$ shown in Fig.~\ref{fig:3D_TDVP}a, the occupation undergoes a slow relaxation with its distribution clinging to the step function form; the difference between the last initially occupied orbital and the first empty one rests on a finite value $\Delta n\lesssim 1$. Such discontinuous occupation spectrum with highly non-ergodic shape, even in the weakly interacting regime, mimics the presence of Fock-space localization. 
We note that the apparent non-ergodic dynamics quenching from DW fermionic states is feasible in the interaction strength in which the CDW pattern is expected to end up thermal (see Fig.~\ref{fig:inhomo_density} and Fig.~\ref{fig:Extended_slow}). This point elucidates the prominent role of the thermal bubbles dimension (initially present in fermion subsystem) on the resolution of this class of dynamical localization. 

To further substantiate such arguments, we examine the asymptotic behavior of the occupation spectrum through time averaged OPDM,
\begin{align} \label{eq:T-OPDM}
\bar{\rho}^f=\frac{1}{\tau}\int_\tau\rho^f(t) dt	,
\end{align}
in the interval $10^5 \leq \tau \leq10^9$, when both the spin and fermion observables, definitely, reach their own steady states. To this end, by substituting for $|\Psi(t) \rangle$ from Eq.~\ref{eq:quench} and using the definitions in Eq.~\ref{eq:gc}, we obtain the exact relation,
\begin{align}{\label{eq:OPDM_eff}}
    \rho^f_{ij}(t)= \frac{1}{\mathcal{N}}\sum_{\{g_i\}} \langle\mathcal{M_{\mathcal{G}}}|e^{it\, H_\mathcal{M}^{\{\bar{g}_i\}}}\,\,c^{\dagger}_ic_j\,\, e^{-it\, H_\mathcal{M}^{\{\bar{g}_j\}}} |\mathcal{M_{\mathcal{G}}}\rangle,
\end{align}
in which only the matter degrees of freedom are involved. Using the above expression makes the ED calculation of $\bar {\rho}^f_{ij}$ strictly straightforward. By performing disorder sampling in Eq.~\ref{eq:OPDM_eff}, and noting that $\rho^f_{ij}(t) \neq \rho^c_{ij}(t)\equiv \langle c^{\dagger}_i c_j\rangle_{|\mathcal{M_{\mathcal{G}}}\rangle, H^{\{\lambda_i\}}_{\mathcal{M}}}$, the corresponding eigenvalues $\langle \bar{n}_\alpha \rangle$ can be captured up to large systems of $L=2N=32$ sites (see Appendix.~\ref{App:matter_OPDM}).

As depicted in Fig.~\ref{fig:OPDM_Fig4}a, in the ergodic phase both the fermionic patterns develop identical gapless occupation spectrum. However, at $\lambda=3.5$ they are notably dissimilar, with the CDW akin to the thermal behavior. 
At the same interaction strength for DW pre-quench state, the average discontinuity of quasi-particle weight $\langle \Delta \bar{n} \rangle$, is observed to be non-ergodic, albeit without any contributions from the single-particle space. The finite-size scaling of $\langle \Delta \bar{n} \rangle$, shown in Fig.~\ref{fig:OPDM_Fig4}b, also support this claim, as it increases with system size and remains finite in the hydrodynamics limit. On the other hand, in the thermal phase (e.g., $\lambda=0.5$ for DW and $\lambda \lesssim 7.5$ for the case of CDW) the discontinuity vanishes linearly with system sizes. Additionally, regardless of both initial fermionic pattern and interaction strength, the resulting asymptotic behavior of the occupation spectrum is well described by the diagonal ensemble corresponding to the Hamiltonian~(\ref{eq:disordered_H}). This is in spite of the fact that $H^{\{\lambda_i\}}_{\mathcal{M}}$ and $H(\lambda,V,0)$ are basically inequivalent and only have similar dynamics.
\begin{figure}[t!]
\centering
\includegraphics[width=1\linewidth]{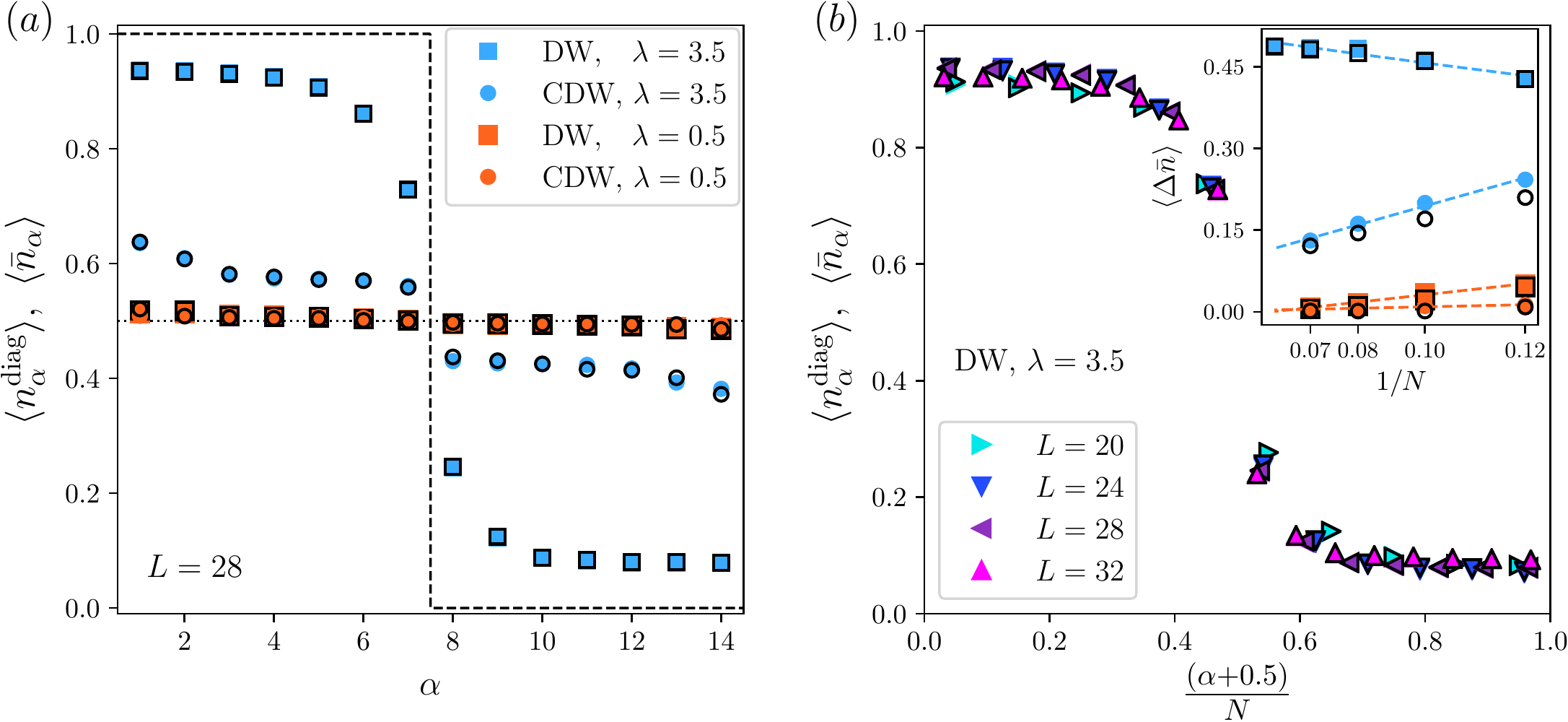}
\caption{(a) Comparison between the steady-state occupation spectrum, quenching from CDW and DW states at two distinctive interaction strength $\lambda=0.5, 3.5$. (b) The same quantity as (a) for a DW state and various length scales up to $L=32$. (inset) Finite-size dependence of the discontinuity $\langle \Delta \bar n \rangle$. Unfilled symbols indicate the diagonal ensemble distribution.}
\label{fig:OPDM_Fig4} 
\end{figure}

Hereby, at moderate interaction strength, a family of pre-quench states $|\psi_0 \rangle_{ref}=|\psi_0^\sigma\rangle\otimes |DW\rangle$ with a discontinuous steady-state occupation spectrum can still serve as the proper \textit{dynamical} reference states~\cite{note:1}. Moreover, divergent collection of states at the same energy density as $|\psi_0 \rangle_{ref}$, but with varied expanse of the insulating blocks would be more tended to draw a continuous, smooth quasi-particle weight. 
Such difference would be despite the fact that all the pre-quench states of the form Eq.~\ref{eq:psi0} 
with the spin product structure, regardless of the their specific fermionic pattern, belong to the infinite temperature ensemble of the original Hamiltonian $H(\lambda,V,0)$.
These findings justify the aforementioned arguments regarding the emergence of dynamical reference states and resolve a pertinent effective temperature in a clean model with the complete failure of the adiabatic continuity. On the other hand, deep in the SBL phase, the asymptotic occupation spectrum of those pre-quench states with dominant thermal clusters (e.g., CDW as an extreme case), similar to that of DW state has a non-vanishing gap(see Appendix.~\ref{App:deep_SBL_OPDM} for more details).

\subsection{Entanglement dynamics
} \label{sec:Sent_dynamics}
\begin{figure}[t!]
\centering
\includegraphics[width=0.57\linewidth]{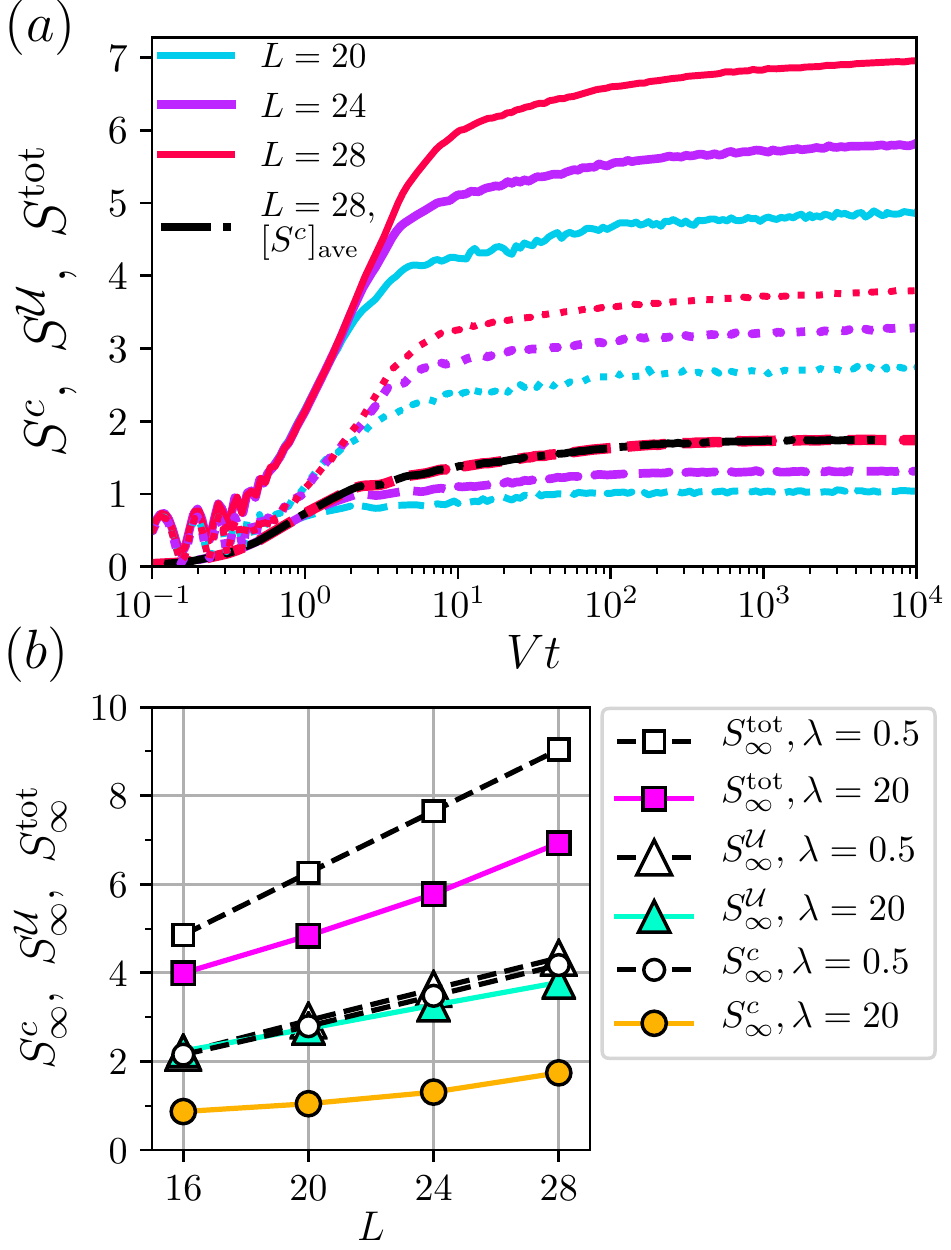}
\caption{
(a) Evolution of total bipartite EE $S^{tot}$ (solid lines) and its projection onto the matter fields $S^{c}$ (dashed lines) and local background charges $S^{\mathcal{U}}$ (dotted lines) at the strong interacting regime $\lambda=20$. The initial state is prepared in $\otimes_{i=1}^{N}|x+\rangle_i \otimes |CDW\rangle$. Dotted-dashed line represents the entanglement dynamics governed by the effective model~(\ref{eq:H_M}) through sampling over $6000$ gauge configurations. (b) Comparison between the scaling of half-cut EE at saturation, for $\lambda=0.5$ and $\lambda=20$, as two extreme cases.}
\label{fig:_qdl_CDW} 
\end{figure}
To further shed light on the non-ergodic dynamics of the system, we address the time evolution of the entanglement entropy (EE) $S^{tot}(t)=-\rho_{A}(t)\,\text{log}_e\,\rho_{A}$ (as an indicator of heating procedure) with $\rho_{A}(t)=\text{Tr}_B|\Psi(t)\rangle\langle \Psi(t)|$ being the reduced density matrix of subsystem $A$. Additionally, we have aimed at evaluating the excess quantity of entanglement in a partially-collapsed state $|\Phi_{\{\nu\}}\rangle$ for a measurement outcome $\{\nu\}$~\cite{Grover:2014}. We can work out the post-measured EE's projected onto the matter $S^{\mathcal{U}}$ and \textit{local} background charges $S^c$, where
\begin{align} \label{eq:QDL}
S^{\bar{\nu}}=\sum_{\{\nu\}}\text{Prob}(\{\nu\})\,S(|\Phi_{\{\nu\}}\rangle),
\end{align}
denotes the EE of the component $\bar{\nu}$ after projecting onto the state of the other one, and $\text{Prob}(\{\nu\})\equiv|\langle \Psi|\Phi_{\{\nu\}}\rangle|^{2}$ is the probability of outcome $\{\nu\}$ given by the Born rule. 
In order to reduce the computational costs of measurement processes, without the loss of generality, we assume the spin subsystem to be initialized in the $x$-polarized state, i.e., $\otimes_{i=1}^{N}|x+\rangle_i \equiv \otimes_{i=1}^{N}|g_i^x+\rangle $. 

For the initial CDW state, even at strong interactions, the bipartite EE $S^{tot}$, shown in Fig.~\ref{fig:_qdl_CDW}a, rapidly saturates to a finite size plateau with strong $L$ dependence. Nonetheless, as depicted in Fig.~\ref{fig:_qdl_CDW}b, by increasing system sizes the corresponding saturation values, $S^{tot}_{\infty}$, deviate from those associated with the delocalized phase which are almost equal to the infinite temperature value $S_{Page}(L)=\frac{1}{2}(L\,{log}\,2-1)$~\cite{Page:1993}, heralding the partial thermalization of the system. The time trace of $S^{\mathcal{U}}$ is also akin to that of $S^{tot}(t)$, yet its asymptotic value coincides with the one corresponding to $S_{Page}(N)$ and obeys volume-law scaling. Indeed, starting from an {\it{arbitrary}} spin product state could not put restriction on the value of the static background charges, resulting in the complete thermalization of $2^N$ gauge fields.
\begin{figure}[t!]
\centering
\includegraphics[width=0.8\linewidth]{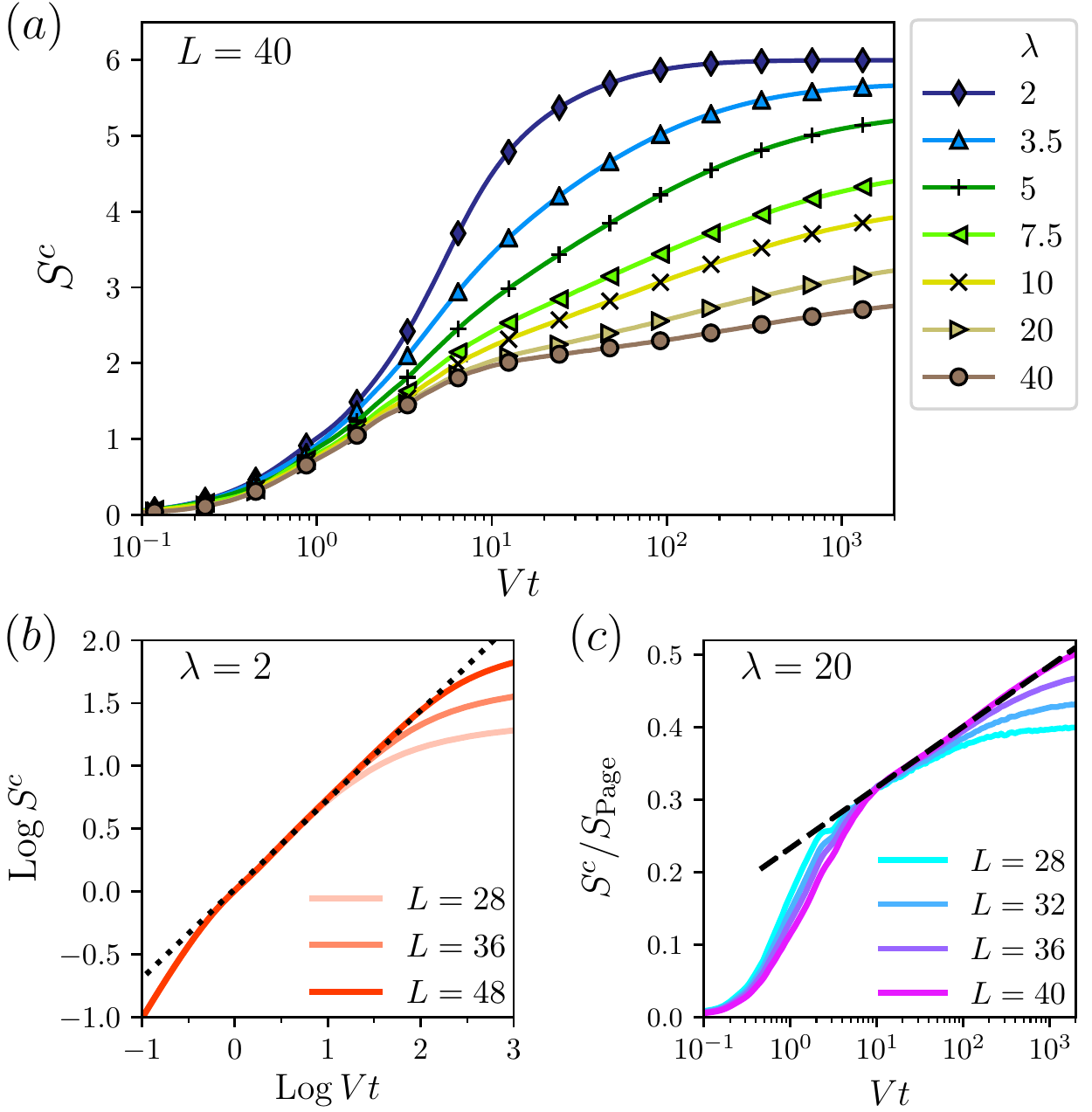}
\caption{(a) Dynamical behavior of $S^c$ and its interaction dependence for $L=40$ using Chebyshev time integration. 
(b) Sub-ballistic growth of $S^c(t)\propto t^{\beta_{\lambda}}$ with $\beta_\lambda \approx0.71$ at $\lambda=2$.
(c) Unbounded logarithmic growth of $S^c$ with system size in the strongly interacting limit.}
\label{fig:sc_subballistic} 
\end{figure}

Conversely, in the strong interacting regime $S^{c}$ possesses a tangibly slow dynamics, which confirms the localization of matter degrees of freedom. However, it is hard to categorize the heating nature of the matter fields through monitoring this quantity for the system sizes accessible by ED. We remark that, while the calculation of $S^{tot}(t)$ and $S^{\mathcal{U}}$ requires the wave function of the whole system, which restricts us to the systems of sizes $L \leq 28$, simulating the evolution of $S^{c}$ could be done more efficiently. 
Indeed, the projective entanglement dynamics of the matter fields governed by the original Hamiltonian $H(\lambda,V,0)$ exactly matches to $[S^{c}(t)]_{ave}$, extracted from the effective Hamiltonian $H_\mathcal{M}^{\{\lambda_i\}}$ after configuration sampling (see dot-dashed line in Fig.~\ref{fig:_qdl_CDW}a). This stems from factorized structure of the spectrum and enables us to capture $S^{c}(t)$ up to the large systems of $L=48$ sites. Moreover, regarding the mentioned equivalence between $S^{c}$ and $[S^c]_{ave}$, the projective entanglement dynamics of the matter fields is independent from the specific choice of initial spin configuration in the pre-quench product state.    

By doing so, in Fig.~\ref{fig:sc_subballistic}a, we have presented the interaction dependence of $S^{c}(t)$, for the same values of $\lambda$ considered in Fig.~\ref{fig:inhomo_density}a. 
This result clearly demonstrate a systematic change from a sub-ballistic growth of $S^c(t)\propto t^{\beta_\lambda}$ for small and moderate interactions to an unbounded logarithmic one in the localized phase. The fit windows extend with system size for both cases, which is indeed expected for the behaviors that hold at the thermodynamic limit. 
In the delocalized phase, the observed sub-ballistic growth in a small fit window at moderate times, shown in Fig.~\ref{fig:sc_subballistic}b, is in analogy with the prescribed power-law decay of imbalance in the regime prior to the transition point (see Fig.~\ref{fig:Extended_slow}). On the other hand, at strong interaction regime, Fig.~\ref{fig:sc_subballistic}c highlights an unbounded logarithmic growth of $S^{c}$, yet without single-particle area-law plateau as expected. It should be noted that owing to the absence of confinement~\cite{Kormos:2016} in the gauge sector of our model, we do not expect to observe the reported extremely slow double logarithmic growth of EE, which has been conjectured to be a dynamical manifestation of self-localization in a confined LGTs~\cite{Brenes:2018}. 
\begin{figure}[t!]
\centering
\includegraphics[width=1.0\linewidth]{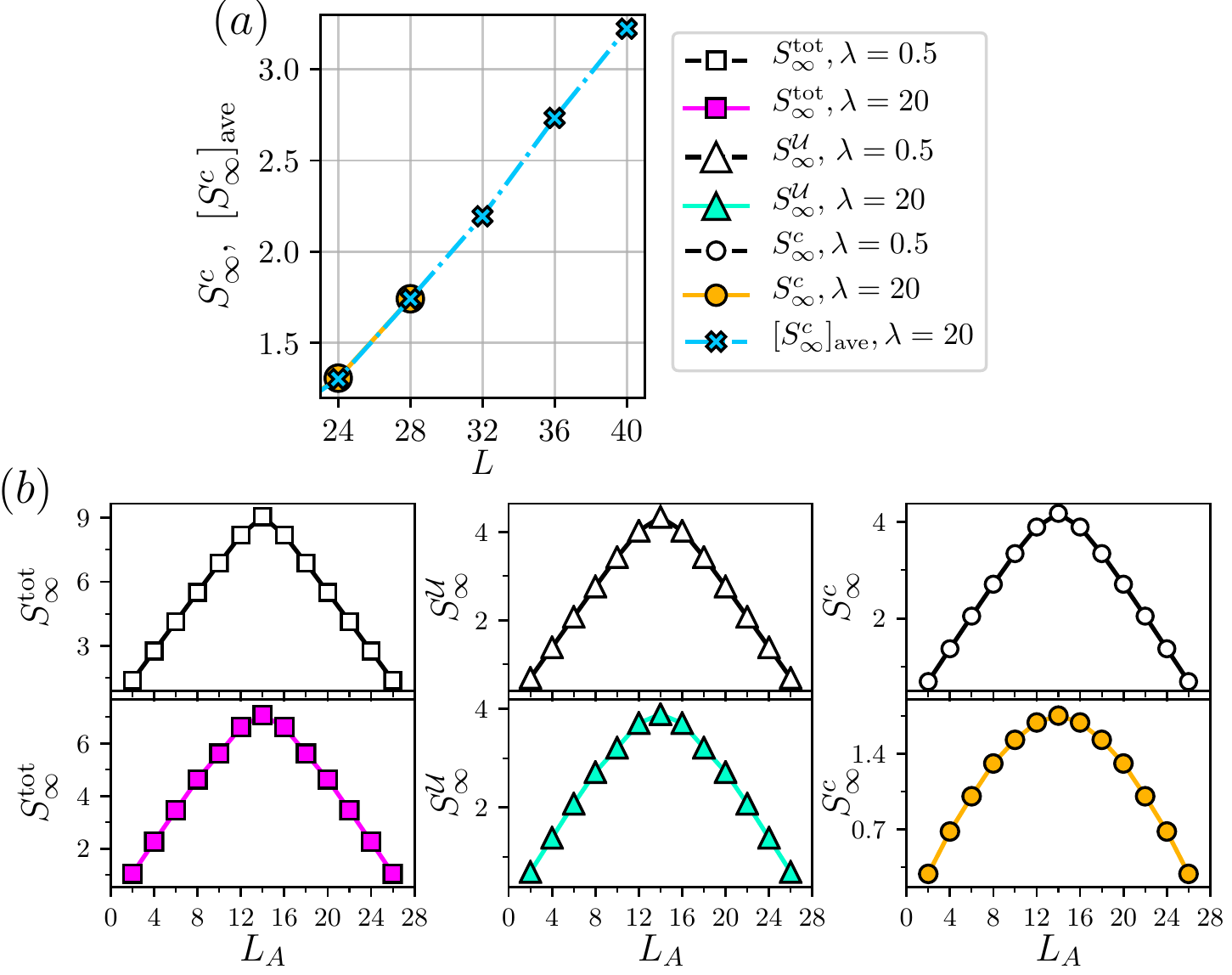}
\caption{(a) Finite-size scaling of saturation bipartite EE at $\lambda=20$. The system initially prepared in CDW fermionic pattern. Note the drastically different scale on the vertical axis compared to Fig.~\ref{fig:_qdl_CDW}b, where $S^c_{\infty}(L)$ depends weakly on system size for $L<24$ and rather remains constant. (b) The volume-law scaling of total and projective saturation EE with varying cutting bond $L_A$. The result are obtained using ED at $L=28$ for $\lambda=0.5$ (top row) and $\lambda=20$ (bottom row).}
\label{fig:QDL_Grover} 
\end{figure}

Moreover, the scaled value of $S^c(t)$, shown in Fig.~\ref{fig:sc_subballistic}c, clearly marks the SBL nature of the matter dynamics, since, despite being slow, considerably reaches around half of its own infinite temperature value even at strong interacting regime. 
Additionally, as depicted in Fig.~\ref{fig:QDL_Grover}a, by considering large length scales $L>24$, we see a noticeable $L$ dependence for $S^c_{\infty}(L)$ with the values extremely larger than those observed deep in on-site disorder cMBL phase; the behavior which is a dynamical diagnostics of SBL phase~\cite{Li:2017}. These atypical asymptotic values can be attributed to the presence of some remnant ergodic regions within the fermionic subsystem, i.e., thermal bubbles, which are left untouched even after measuring the primary ergodic components $\mathcal{U}_i$s. Thus, our results affirmatively confirm the occurrence of emergent SBL.


\begin{figure}[t!]
\centering
\includegraphics[width=0.715\linewidth]{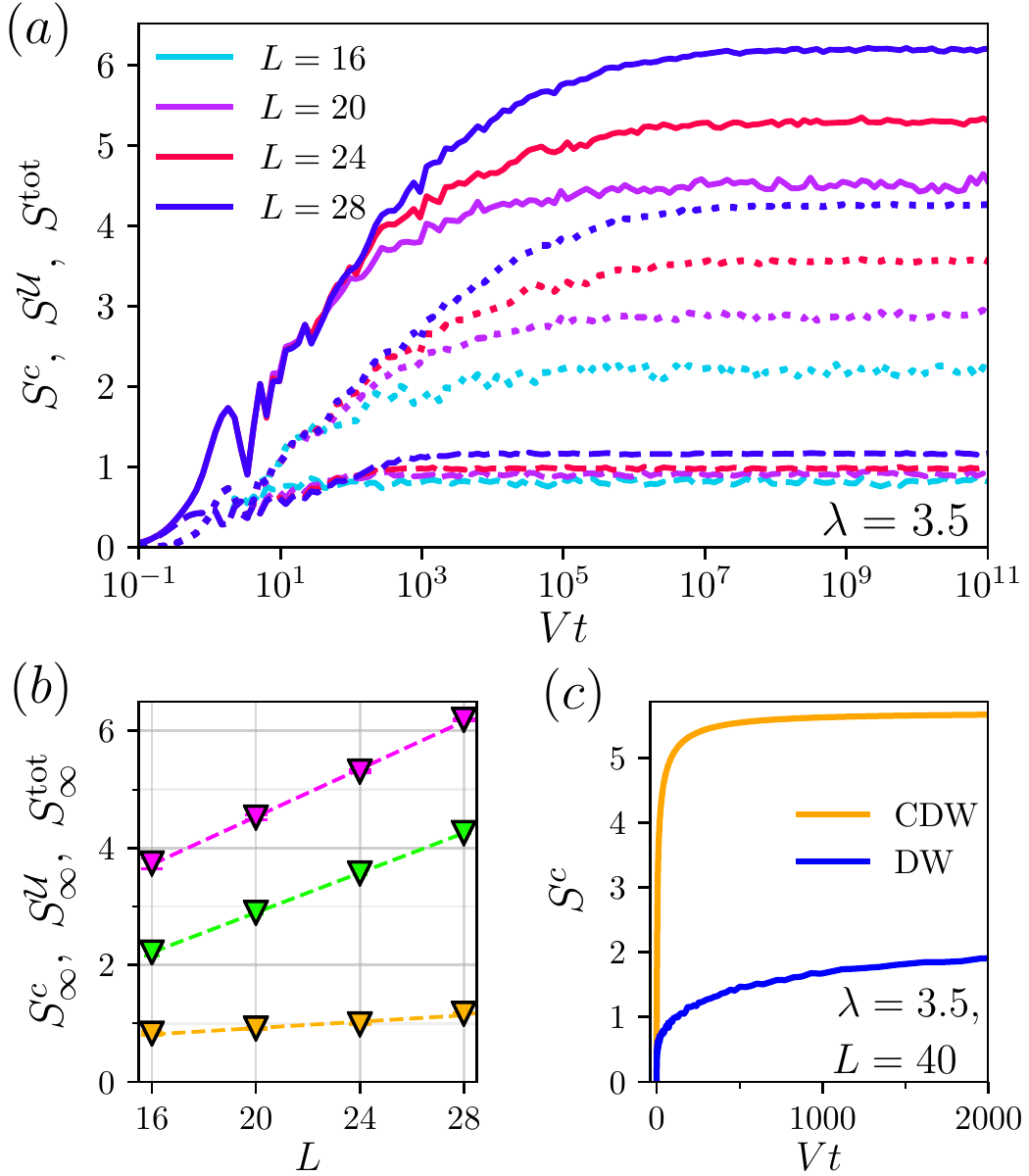}
\caption{(a) The dynamics of $S^{tot}$, $S^{\mathcal{U}}$ and $S^{c}$ (denoted by solid, dotted and dashed lines, respectively) at $\lambda=3.5$, starting from $\otimes_{i=1}^{N}|x+\rangle_i \otimes |DW\rangle$. (b) Finite size scaling of the late time 
values of the quantities in (a) averaged on time period $Vt\geq10^{11}$. (c) Comparison between the matter entanglement dynamics of the CDW and DW pre-quench states.}
\label{fig:qdl_DW} 
\end{figure}

The coexistence of such localized and ergodic subsystems  
along with the extensive scaling of $S^{tot}_{\infty}$, 
is reminiscence of the concept of QDL~\cite{Grover:2014}.
Recently, this behavior has been found in a $\mathbb{Z}_2$ gauge invariant model~\cite{Smith:2017-b}, wherein the bipartite EE reforms from a volume-law to an area-law after a projective local measurement of gauge fields. Such partial measurement uncovers hidden locality within Anderson localized effective Hamiltonian dynamics of the matter fields.
Conversely, we are dealing with a $\mathbb{Z}_2$ LGT in which the integrated out model possesses the emergent SBL whose single particle dynamics is completely ergodic, and self-localization occurs in the absence of locality. 
Thereby, the coexistence of disentangled nonergodic and ergodic degrees of freedom in our model can not be monitored by area/volume-law scaling of the saturation post-measured EE.
In order to support this claim, we study the scaling of overall and post-measured von-Neumann EEs with cutting bond as a formal diagnostic of the QDL phase~\cite{Garrison:2016}. Figure.~\ref{fig:QDL_Grover}b typifies the long-time limit of all three entanglement measures which crucially experience the same extensive scaling both in the thermalized regime (top row) and deep in the self- localized one (bottom row). 
The overall presented line of reasoning suggests that
our $\mathbb{Z}_2$ lattice gauge model, despite having disentangled components, exceeds the standard established phenomenology of QDL.

In view of the SBL nature of the effective Hamitonian~\ref{eq:disordered_H}, the dynamics is expected to show quantitative difference from both AL and cMBL, associated with the physics of thermal bubbles. To further resolve this difference, we now turn to the DW initial state. As shown in Fig.~\ref{fig:qdl_DW}, $S^{tot}$ and $S^{\mathcal{U}}$ have the same time trace and display volume law scaling at the late time as well. For small system sizes and moderate interaction strength, the scaling of $S^c_{\infty}(L)$, shown in Fig.~\ref{fig:_qdl_CDW}b, has a qualitatively similar behavior to that of CDW deep in the SBL phase. Indeed, given the notion of dynamical reference state, one expects the heating procedure of the matter fields starting from DW pattern to be much slower than the one corresponded to CDW pre-quench state (see Fig.~\ref{fig:_qdl_CDW}c); the behavior which can be appreciated from the smallest possible amount of initial thermal clusters within the DW pattern.


\subsection{Dynamics of the spin subsystem}
\label{sec:DOTSS}

Up to now, we have argued that a topologically nontrivial spin chain coupled to an interacting, topologically trivial fermionic layer can lead to the new class of disentangled liquids containing non-ergodic and fully thermalized spices. Here we aim at tackling whether such proximity modifies the topological essence of the fermion-cluster model.
It is known that the cMBL can play a crucial role in the protection of quantum order in out-of-equilibrium many-body states~\cite{Huse:2013,Chandran:2014,Bahri:2015,Potter:2015,Yao:2015}.
In this respect, a rewarding question is what are the chances to employ the dynamical localization of the proximate fermions in order to protect the topological edge modes of the cluster spin chain which itself is utterly thermalized? 

\begin{figure}[t!]
\centering
\includegraphics[width=1.0\linewidth]{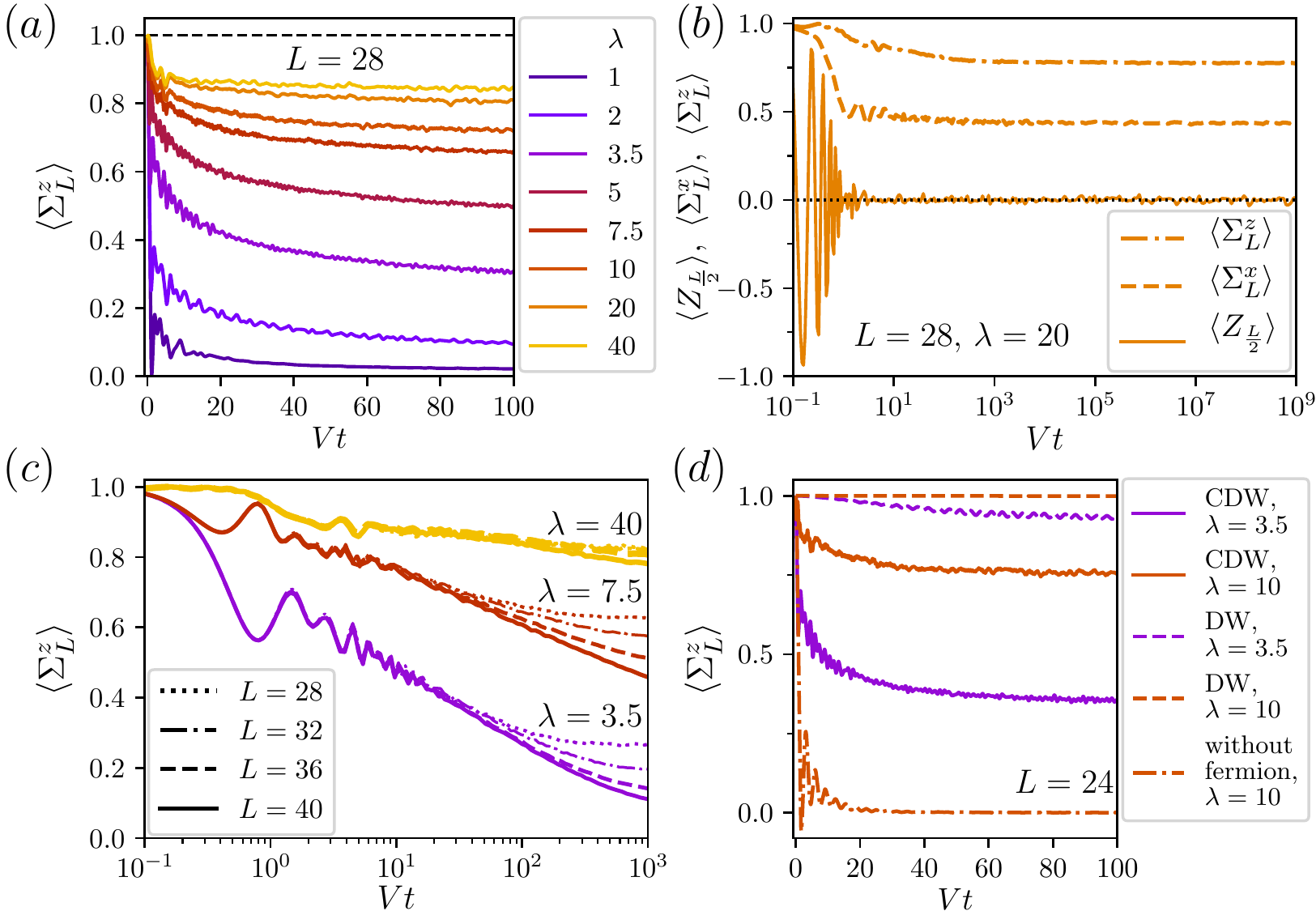}
\caption{(a) Dynamical behavior of the edge spin expectation value $\langle \Sigma^z_L \rangle$ for different interaction strengths $\lambda$, starting from infinite temperature state $\otimes_{i=1}^{N}|z+\rangle_i \otimes |CDW\rangle$. (b) Demonstration of the quantum nature of the long-lived edge spins in comparison with rapid lose of initial information stored in the hot bulk modes.
(c) Decay of the edge mode with system size at different $\lambda$.
(d) The effect of preparatory fermionic pattern on the robustness of topological edge states.}
\label{fig:EDGE} 
\end{figure}

It should be noted that deep in the SBL phase of fermions, i.e., $\lambda \gg V$, puts the spin subsystem in the topologically non-trivial regime of the Hamiltonian~(\ref{eq:H_cf}). This further motivates us to investigate the remarkable interplay between (self-) localization of $f$-fermions and its influence on protection of topological edge modes ${\Sigma}_{L(R)}$ in terms of the original spins. 
With a huge number of correlated, ergodic bulk degrees of freedom and their strong coupling to the edge modes, at the first sight, one might expect the quantum information initially stored in ${\Sigma}_{L(R)}$ to irreversibly disperse due to scattering with thermal bulk excitations. 
Yet, there exists an exact equivalence,
between the temporal autocorrelation functions of the soft edge modes and those corresponding to the parity operators of fermions at the edges, i.e.
\begin{align} \label{eq:proximity_edge}
\langle \Sigma^z_{L(R)}(t) \Sigma^z_{L(R)}(0)\rangle_{|\Psi_0 \rangle}=& \langle (-1)^{n^f_{1(L)}(t)+n^f_{1(L)}(0)} \rangle_{|\Phi_0 \rangle}, \nonumber \\ 
\langle \Sigma^x_{L(R)}(t) \Sigma^x_{L(R)}(0) \rangle_{|\Psi_0 \rangle}=& \langle (-1)^{n^f_{2(L-1)}(t)+n^f_{2(L-1)}(0)} \rangle_{|\Phi_0 \rangle}.
\end{align}
 The pre-quench states $|\Phi_0 \rangle$ and $|\Psi_0 \rangle$ with the form of Eq.~\ref{eq:psi0} have the same fermionic pattern, even though in general, their spin parts could be different product states. The above equality then holds iff $|\Psi_0 \rangle$ exhibits a non-zero expectation value for the edge operators. 
A dramatic consequence of the dynamical equivalence~(\ref{eq:proximity_edge}) is that the localization of fermions, which on its own originates directly from the initial entanglement structure of the spin part, can dynamically leak to the spin edge zero modes. Hence, the self-localization of proximate fermions make the edge modes to behave as dynamically decoupled degrees of freedom from the ergodic bulk. 

Such \textit{proximity-induced self-localization} is
reminiscent of the situation giving rise to the cMBL proximity effect
~\cite{Nandkishore:2015,Hyatt:2017,Marino:2018},
the phenomena in which localization can be
induced in the ergodic bath through coupling to a cMBL system with comparable number of degrees of freedom. 
However, in our case the mechanism behind the induction of self-localization into edge zero modes comes purely from dynamical origin, and hence completely differs from the one suggested by Ref.~\onlinecite{Nandkishore:2015}.

From another point of view, while the (effective) Hamiltonian of ($c$)$f$-fermion degrees of freedom has no topological character, the dynamics of its edge operators is equivalent to those of topologically non-trivial cluster chain.  
This is an interesting variant of the concept of "bulk topological proximity effect"~\cite{Hsieh:2016} wherein the low-lying energy sector of an ordinary system can itself exhibit nontrivial topological character by proximity of 
a topologically nontrivial one. However, here, topological properties are induced solely through dynamics,in the sense that the fermions at the edges of a topologically trivial system behave like the edge modes of the cluster chain which has SPT character.


In the first set of calculations we deliver the findings, provided by ED, regrading the dynamical proximity effect on stabilizing the soft MZMs. 
We again note that the right hand side of Eq.~\ref{eq:proximity_edge} is invariant for any initial spin product state with an arbitrary high energy density. Thereby, the dynamics of the edge modes is independent of the specific choice of initial spin product configuration as long as it exhibits a non-zero value for the edge operators. The results of Fig.~\ref{fig:EDGE}a for $L=28$ elucidate that by systematically increasing the interaction strength, the relaxation of $\langle\Sigma^{Z}_L\rangle$ slows down and approaches a non-zero stationary value at the long times. This can be interpreted as a first clue that the initial information encoded in the edge modes persists owning to the localization of proximate fermions. On the other hand, even deep in the localized phase of fermions, the expectation value of spin at the middle of the chain, shown in Fig~\ref{fig:EDGE}b, decay rapidly to zero which is consistent with the ergodic nature of bulk spin modes. At the same time, all components of the topological edge state can survive for the longest time, conveying that these long-lived edge spins can indeed store a quantum bit of information. 

To explore whether this is a robust effect that persists in the thermodynamic limit, in Fig.~\ref{fig:EDGE}c, we examine how this behavior depends on system size. On approaching the fermionic localization transition, the relaxation of the topological edge state at intermediate time displays a logarithmic decay subsequent to an initial sinusoidal damping; the behavior that is clearest for larger system size $L=40$. Then, it eventually saturates to its finite size thermal value, signaling the complete loss of initial information. However, this is not the case for the interactions strengths deep in localized phase, as the corresponding edge dynamics in this regime shows a strong non-ergodic behavior.

\begin{figure}[t!]
\centering
\includegraphics[width=0.8\linewidth]{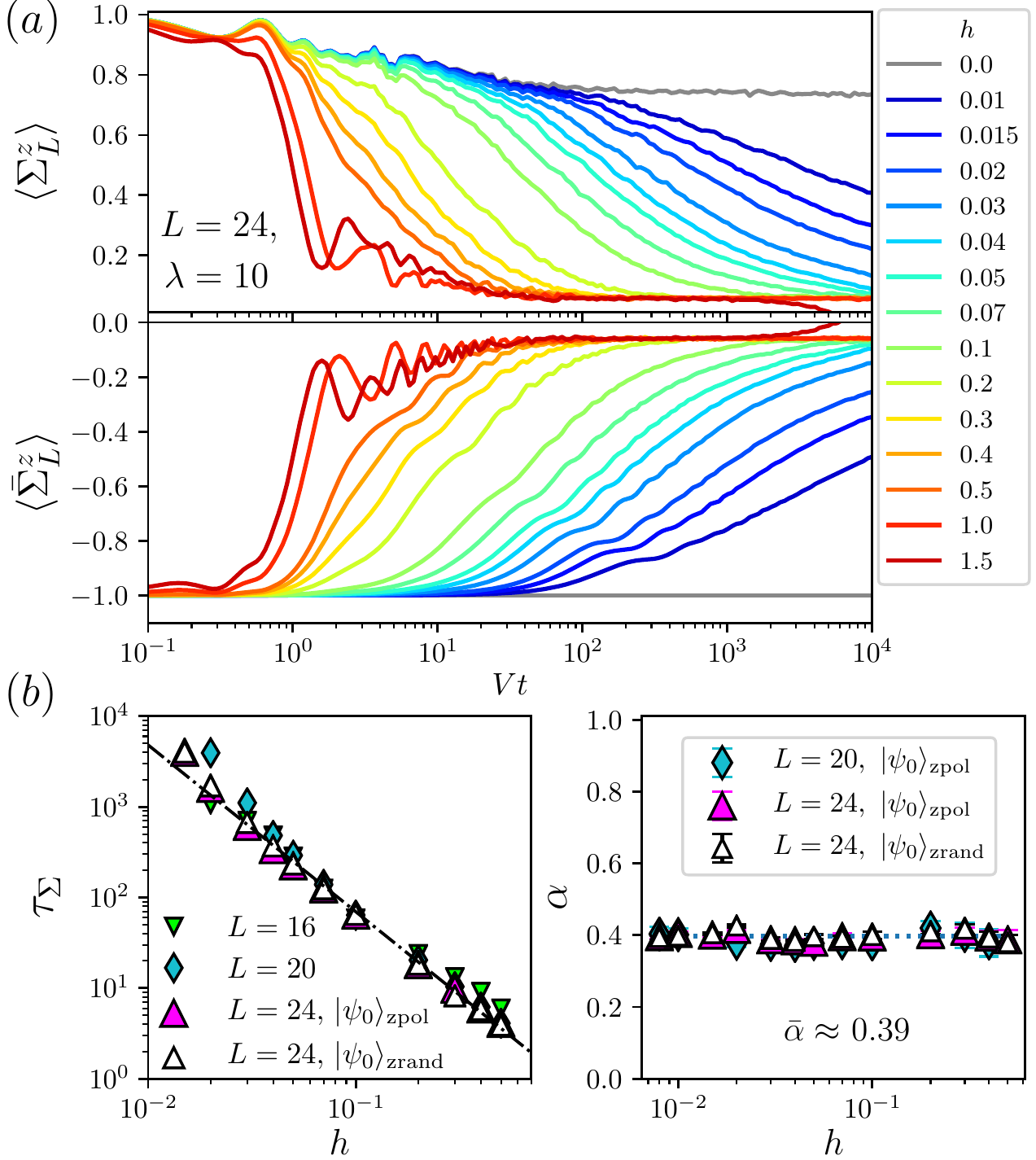}
\caption{(a) Stretched slow decay of the spin edge qubits (top panel) and the strong zero modes (bottom panel) at $\lambda=10$ and $L=24$ for different values of external field. (b) The characteristic parameters of the stretched slow decay of $\langle\Sigma^{z}_{L}\rangle$  for different system sizes. The results are extracted from two high temperature product initial states $|\psi_0 \rangle_\text{zpol}$ and $|\psi_0 \rangle_\text{zrand}$. Left and right panels represent the edge coherence time and stretching exponent, respectively.}
\label{fig:edge_hx} 
\end{figure}
\begin{figure*}[t!]
\centering
\includegraphics[width=.85\linewidth]{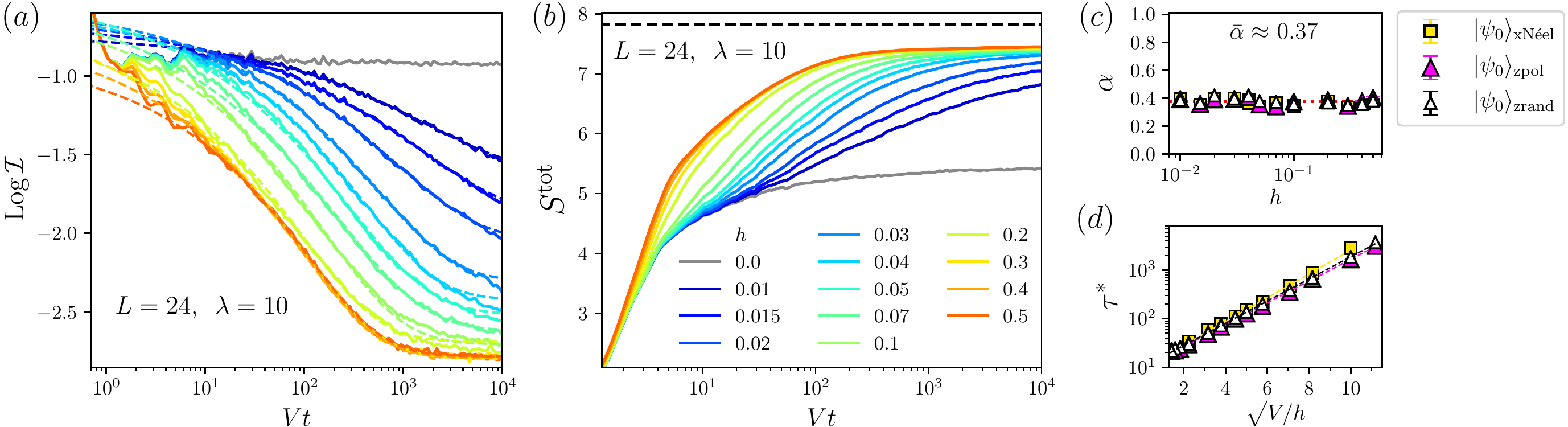}
\caption{ (a) Stretched slow decay of imbalance for different values of $h$. The dashed lines indicate the stretched fitting in Log-Log plot (the fitting function is $\mathcal{I}^{sat}_{h=0} e^{\Gamma t^{\alpha}}$, where $\mathcal{I}^{sat}_{h=0}$ is the sturation value of imbalance in the absence of external field). (b) Growth of von-Neumann entropy for the same parameters in (a). The black dashed line designates the infinite-temperature value $S_{Page}$. (c) The universal behavior of the  stretching exponent $\alpha$, corresponding to the decay of imbalance for various pure initial states (at the middle of the energy band). (d) The exponential scaling of the thermalization time $\tau^{*}$ extracted from entanglement dynamics in (b).}
\label{fig:imb_sent} 
\end{figure*}

In order to confirm the necessity of the presence of proximate fermions on preserving the topological facet of the ergodic subsystem, in Fig.~\ref{fig:EDGE}d, we also consider the dynamics of the soft edge modes in the absence of fermionic degrees of freedom, governed by the cluster Hamiltonian~(\ref{eq:H_cluster}) with $L=24$ spins. Without the fermionic degrees of freedom, and thus no dynamical localization, even at strong interactions, the edge modes rapidly go out of order. In contrast, for the same interaction strengths (and also the same lattice site) the soft edge modes, evolved under the cluster-fermion Hamiltonian, have an infinite lifetime which asserts that the protection of edge modes is solely due to localization of proximate fermions. 

Finally, by considering DW and CDW pattern, we investigate the effect of initial fermionic thermal bubble on controlling the dynamics of the edge modes. Even at moderate interaction, the dynamics of the edge modes starting from DW fermionic configuration has more tendency toward localization than the one initialized in the CDW pattern. This behavior can be interpreted as evidence for the role of initial thermal bubbles in the potency of fermionic self-localization through SBL mechanism, which is now encoded in the robustness of topological edge states, as expected by unraveling Eq.~(\ref{eq:proximity_edge}).

\section{Robustness of statistical bubble self-localization under generic perturbations}\label{sec:bath-like}

So far, we have identified proximity-induced self-localization, 
as a by-product of dynamical localization of fermions. Here we examine how fragile are these phenomena in the face of more generic perturbations, e.g., an external filed with the form of $h\sum_j X_{j}$ which explicitly breaks the local $\mathbb{Z}_2$ invariance~(\ref{eq:gauge-symmetry}) but at the same time respects to the $\mathbb{Z}_2 \times \mathbb{Z}_2$ symmetry generated by $P_{e/o}$. 
As the ramification of adding such terms, the spectrum of the Hamiltonian $H(\lambda,V,h)$ can no longer be factorized into two well-separable sectors (note that the effective Hamiltonian~(\ref{eq:H_gc}) in terms of $\hat{g}$-spins and $c$-fermions is still valid), and the dynamics of previously disentangled matter and gauge fields becomes inextricably intertwined. 
From this perspective, the ergodic $\hat{g}$-spins can collectively serve as a small bath for the previously self-localized $c$-fermions, which leads to the inevitable thermalization of entire system characterized by a stretched exponential slow relaxation. 
Indeed, for those values of interaction strengths $\lambda$, corresponded to the SBL phase of fermions at $h=0$, turning on gauge spoiling perturbations would result a concomitant change of time correlators to a stretched exponential form.  

This should be noted that in the presence of external fields, the dynamics of fermions (and also the edge modes) would no longer be invariant under quenching from different spin product configurations. Hence, in order to capture the infinite temperature response of the system, one oughts to restrict dynamics to the pre-quench states within infinite temperature ensemble of $H(\lambda,V,h)$. From later on, we fix the CDW pattern for the fermionic part of pre-quench state and let the spin part to be initialized in three different product states: $z$-polarized and N\'eel state in $x$ direction (as two translationally invariant states) as well as an inhomogeneous random product states in $z$ direction (these states are denoted by $|\Psi \rangle_\text{zpol}$, $|\Psi \rangle_\text{xN\'eel}$ and $|\Psi \rangle_\text{zrand}$, respectively).

\subsection{Long-time coherence of topological qubits}
Since the commutation relation~(\ref{eq:comm_strogn_modes}) does not hold for any non zero value of $h$, the term ``strong'' can not be further attached to the edge modes of the composite system $\bar{\Sigma}^{z}_{L(R)}$. 
However, Fig.~\ref{fig:edge_hx} signifies that the topological edge modes can be stabilized substantially as the relaxation of both $\langle \bar{\Sigma}^{z}_{L}\rangle$ and $\langle\Sigma^{z}_{L}\rangle$ 
still display a lingering signature of localization  characterized by a stretched exponential decay of the form $e^{-(t/\tau_\Sigma)^\alpha}$ ($\alpha$ and $\tau_\Sigma$ denote the stretching exponent and the lifetime of the topological edge modes, respectively). 
The life time of the spin edge mode is approximately independent of the system size and manifests an apparent power law decay by progressively increasing $h$ in the regime $h\lesssim V\ll\lambda$ (see Fig.~\ref{fig:edge_hx}b). 
Besides, the exponent $\alpha$ possesses the universal value ($\approx0.39$ for the case $\lambda=10$) confirmed for different system sizes and generic pure initial states with and without translational invariance~\cite{footnote}. Interestingly, the same behavior (albeit with different universal value of $\alpha$) have been also reported for the relaxation of edge modes in topologically non-trivial cMBL system coupled to the Markovian dissipative 
bath~\cite{Carmele:2015}.

\subsection{The exponentially slow heating and dynamics of fermionic bulk observables}
The dynamics of the imbalance, depicted in Figs.~\ref{fig:imb_sent}a,c, indicates that the fermionic observables have also a stretched slow decay with the universal behavior akin to that of the edge modes. 
This residual slow approach to eventual thermal equilibrium clearly manifests that the $\hat{g}$ components are acting collectively as an ergodic bath which tends to destroy the dynamical SBL phase of the fermions. 
 
To further probe the heating nature of composite system in the manner which is not relying upon a specific choice of operators, we gate the dynamics of the half-cut EE. By increasing the strength of gauge spoiling term, the dynamics of $S^{tot}$ changes its own behavior from a slow growth for the case of massive $\hat{g}$-particles (i.e., $h \lesssim V$) to a fast one at $h\approx V$, where both $c$-fermions and $\hat{g}$-spins have nearly identical masses (see Fig.~\ref{fig:imb_sent}b). Specially, for the case of $h>0.03$ the EE eventually reaches to a value very close to $S_{Page}(L)$ within the time scales accessible by our numerics. This feature also ensures the immunity of our results against finite size effects. In order to explore the heating time associated with such slow growth, we consider the thermalization time scale $\tau^{*}$~\cite{Machado:2017}, at which half-cut EE is halfway from the late time value for $h=0$ to its late time infinite temperature plateau. We found that $\tau^{*}$, shown in Fig.~\ref{fig:imb_sent}d, has an exponential dependence on the perturbing coupling for a variety of different initial states (including inhomogeneous as well as translationally invariant states), conveying the exponential slow heating of the system.  Such coarse-grained behavior has been previously observed in the relaxation of cMBL system weakly coupled to either a dissipative environment~\cite{Fischer:2016,Medvedyeva:2016,Levi:2016,Everest:2017} or an internal subsystem with bath-like structure~\cite{Bordia:2016,Luschen:2017-b}, as well as spin-glasses~\cite{Chamberlin:1984}, fractal structures~\cite{Even:1984} and kinetically constrained models~\cite{Ritort:2003}.

\section{Summary and Discussions}\label{sec:discuss}

In summery, we introduce a clean $ \mathbb{Z}_2$ gauge invariant model in which the matter components are dynamically localized \textit{due to and not despite} of (inter and intra particle) interactions. While the self-localized matter fields are embedded throughout the fully thermalizing gauge fields, they still show many dynamical diagnostics of cMBL. This suggests that ergodicity (and translational symmetry) can be dynamically broken in a generic model where only pure  many-body effects are at play and none of the well-known phenomenological descriptions of the cMBL are valid. Through varying the entanglement structure of the initial states and the strength of different types of interaction and external field, the dynamics captures a rich phase structure containing extended ergodic, extended slow, localized and stretched slow behavior. 



Through comparing the OPDM (and also entanglement) dynamics of two extreme cases, i.e., CDW and DW preparatory fermionic arrangement, we characterized the influence of the initial fermionic thermal bubbles on the persistency of the emergent localization.  In this case, an effective temperature is provided by tuning the length of initial thermal bubbles relative to DW pre-quench state. Such perfect insulating block serves as a dynamical reference state with a (zero-temperature) discontinuity in its asymptotic OPDM spectrum. Despite the absence of adiabatic connection to a single-particle insulator, this Fermi-liquid like picture is consistent with the physics of bubble neck eigenstates in the SBL mechanism.

At moderate interactions (equivalent to moderate self-disorder strengths), by varying the extent of the  insulating blocks 
we established a qualitative change in matter's dynamics, from SBL for the DW pattern, to the extended slow dynamics for the CDW pre-quench state. It should be noted that whereas the observed anomalous extended slow behavior is akin to those studies of conventional MBL (with both true-random~\cite{Agarwal:2015,Herrera:2015,Luitz:2016,Znidaric:2016,Lev:2015,Khemani:2017,Doggen:2018} and quasiperiodic~\cite{Bordia:2017,Luschen:2017,Lev:2017,Lee:2017,Khemani:2017} \textit{single-particle} potentials), the underling mechanisms behind this anomalous behavior is different from the previous ones. Recent studies~\cite{Luschen:2017} on quasiperiodic systems with long-range correlated disorder have unveiled that the Griffiths picture~\cite{Vosk:2015,Potter:2015,Gopalakrishnan:2016} does not faithfully warrant the anomalous relaxation. Rather, such extended slow thermalization is attributed to some spatially atypical regions in the initially chosen states from the infinite temperature ensemble~\cite{Luitz:2016,Luschen:2017}. Yet it has been argued that identical trends also hold true even for the relaxation of translationally invariant, pure initial states~\cite{Lev:2017} (e.g. CDW or Neel state), which is speculated to stem from atypical transition rates (with broadly distributed hopping times) between localized \textit{single-particle} states~\cite{Luitz:2016-b,Lev:2017}. 
The applicability of all the outlined mechanisms thereupon, in some sense, is associated with the presence of external disorder, a determinable single-particle localization length (in the non-interacting framework), and hence the validity of LIOM picture. Conversely, in our setting there is no requirement of both randomness (neither in the initial state nor in the Hamiltonian) and (self-) localization of single particle states. On this account, our study calls for an additional gadget to expound on the extended slow dynamic, observed in our clean model. 

Moreover, we unveil proximity-induced self-localization, which prevents bulk spin excitations from carrying quantum information away from topological MZMs. Whether they can be manipulated coherently and how one can retrieve the quantum coherence of these MZMs through spin-echo procedures, can be subject of further investigations. Our results are comparable to those of Ref.~\onlinecite{Bahri:2015} in which the edge modes of the dirty cluster chain was decoupled from the hot bulk due to the presence of LIOMs in the bulk, even though, here the protection of edge spins is achived in a completely ergodic subsystem of a clean model through proximity-induced self-localization in the absence of LIOMs. Looking forward, the protection of zero edge modes in an ergodic system with completely extended bulk modes might make our setting a viable platform for the implementation of recent proposals on storing and shuttling the SPT qubits~\cite{Yao:2015}.

Additionally,  we show that the cluster-fermion Hamiltonian posses physically addressable edge quibits $\bar{\Sigma_{L(R)}}$ whose  appearance in a generic model with an incredibly complex dynamics is rather surprising.
In the typical studies of {\it{integrable}} systems, manifestation of such modes with a lasting coherence time up to exponential finite size corrections is proved in examples like transverse-field Ising chain~\cite{Kemp:2017} (identical to the non-interacting Kitaev chain) as well as the interacting XYZ model~\cite{Fendley:2016}. By inclusion of integrability breaking terms, in the best case scenario some perturbative arguments~\cite{Kemp:2017,Abanin:2017,Else:2017,Else:2017-b}, namely ADHH theorem~\cite{Abanin:2017}, offer {\it{prethermal}} zero modes with a life-time bounded below by a nearly exponential function of some local energy scales. Unlike the above reported schemes, according to the exact commutation relation~(\ref{eq:comm_strogn_modes}), the spectrum of the \textit{non-integrable} model~(\ref{eq:H_M}) 
is entirely paired, assuring an infinite life-time of $\bar{\Sigma}_{L(R)}$ at arbitrary high energy densities.

Finally, we investigated the robustness of dynamical SBL under generic gauge-spoiling perturbations and inferred the stretched exponential relaxation toward the eventual thermal equilibrium. Impressively, this behavior is the one observed in the cMBL systems coupled to either a quantum bath-like structure (with few degrees of freedom )~\cite{Bordia:2016,Luschen:2017-b} or a classical dissipative bath with negligible back action~\cite{Carmele:2015,Fischer:2016,Medvedyeva:2016,Levi:2016,Everest:2017}. For the latter case, such coarse-grained response is justified through an effective classical rate equation based on the validity of the LIOM picture~\cite{Fischer:2016}. In contrast, we observed qualitatively the same response in the model with complete failure of LIOM phenomenology. Moreover, the presence of such behavior in a $\mathbb{Z}_2$ gauge invariant model (subjected to a gauge spoiling field) together with the recent progress on realizability of LGT dynamics~\cite{Martinez:2016} can construct a promising platform for investigation of the interplay between localization and bath-like structures, which is a topic of growing interest~\cite{Luschen:2017-b,Abadal:2018}.

\section{Acknowledgements}
We gratefully acknowledge 
M. Dalmonte, P. Fendley, M. Heyl, S. A. Jafari, S. S. Jahromi, Y. Javanmard, V. Khemani, C. Laumann, A. Vaezi and M. \v{Z}nidari\v{c} for many enlightening and insightful discussions.
Special thanks is devoted to Tal{\'i}a L.~M. Lezama for valuable discussions and comments
on the intial version of manuscript.
The authors would like to thank Sharif University of Technology for financial supports under grant No. G960208, and CPU time from LSPR, Sputtering and Cosmo HPC machines. Our numerical schemes are based on the PETSc~\cite{petsc-user-ref,petsc-efficient}, SLEPc~\cite{Hernandez:2005} and open source MPS libraries~\cite{Wall:2015,Jaschke:2018}.
\appendix

\section{Strong edge zero modes}\label{App:MZM}
In this section we explicitly construct the strong edge zero modes of the Hamiltonian~\ref{eq:H_M}. 
In this respect, we use the duality mapping,

\begin{align}\label{eq:dual_tu}
 \left\{
 \begin{array}{ll}
    \hat{g}^{z}_{2i}&=\tau^{z}_{2i}\prod_{k\leqslant i} \tau^{x}_{2k-1} ,\\ \\
    \hat{g}^{z}_{2i-1}&=\tau^{z}_{2i-1}\prod_{k\geqslant i} \tau^{x}_{2k},
 \end{array}
 \quad, \,\,\,\, \hat{g}^x_{i}=\tau^{x}_i,
\right.
\end{align}
in which $\tau^{\alpha}_i$ operators obey spin-1/2 Pauli algebra and commute with the $c_i$ fermions. The Hamiltonian in the $\tau$-picture again takes the form of the cluster Ising chain coupled to the interacting fermions,
\begin{align}
 H(\lambda,V,0)&=
    \lambda \sum_{i}  n_{i-1} \mathcal{W}^{\tau}_i\, n_{i+1} \nonumber \\
    &+ V \sum_{\langle i j \rangle} c^{\dagger}_{i} c_{i+1}\,,
\end{align}
where $\mathcal{W}^{\tau}_i = \tau^z_{i-1} \tau^x_i \tau^z_{i+1} $. However, due to the absence of $\tau^x_i \tau^x_{i+1}$ term, the edge operators in terms of the $\tau$-spin are the constants of motion. Hence, one can define the strong edge zero modes of the composite system at the left end (and similarly for the right one) as, 
\begin{align}
    \bar{\Sigma}^z_L&= \Sigma^{\tau^z}_L= \tau^z_{1}= g^z_1 \hat{P}_{e}= (-1)^{n_1} \Sigma^z_L , \nonumber \\ 
    \bar{\Sigma}^x_L&= \Sigma^{\tau^x}_L= \tau^x_{1} \tau^z_{2}= g^z_2= (-1)^{n_2} \Sigma^x_L,\\ 
    \bar{\Sigma}^y_L&= \Sigma^{\tau^y}_L= \tau^y_{1} \tau^z_{2}=-i g^z_1 g^z_2 \hat{P}_{e}= (-1)^{n_1+n_2} \Sigma^y_L, \nonumber
\end{align}
which obey the Pauli algebra and exactly commute with $H(\lambda,V,0)$. 
   
 
\section{Derivation of the OPDM in terms of the matter fields}\label{App:matter_OPDM}

In order to prove the exact relation Eq.~\ref{eq:OPDM_eff} in the main text, we first consider the OPDM in terms of $f$-fermions and substitute for $f_i$ from the definition of gauge invariant fermions which results,
\begin{align}
    \rho^f_{ij}(t)  =\langle\Psi_0| e^{itH} \ \hat{g}^x_i c^{\dagger}_i \; \hat{g}^x_j c_j \ e^{itH} |\Psi_0\rangle .
\end{align}
Since the dynamics of the matter degrees of freedom is invariant under the specific choice of initial spin product configurations, without the loss of generality, we let the spin part to be initialized in $x$-polarized state that means $|\Psi_0\rangle=|\mathcal{G}_0\rangle \otimes|\mathcal{M}_{\mathcal{G}_0}\rangle$, where $|\mathcal{G}_0\rangle=\otimes_{i=1}^{N}|x+\rangle_i \equiv \otimes_{i=1}^{N}|g_i^x+\rangle$.

Now, by utilizing the anti-commutation relation between $\hat{g}^x_i$ and $\hat{g}^z_i$ and noting that $\hat{g}^x_{i}=X_i$, the exact representation of OPDM in the language of $c$-fermions reads as follows, 
\begin{align} \label{OPDM_psi_g}
    \rho^f_{ij}(t)=\frac{1}{\mathcal{N}}\sum_{\{g_i\}} \langle\mathcal{M}_{\mathcal{G}_0}|e^{it\, H_\mathcal{M}^{\{\bar{g}_i\}}}\,\,c^{\dagger}_ic_j\,\, e^{-it\, H_\mathcal{M}^{\{\bar{g}_j\}}}|\mathcal{M}_{\mathcal{G}_0}\rangle.
\end{align}
Here, the sign of the charges identified by $\{\bar{g}_i\}$ is opposite to those appearing in Eq.~\ref{eq:H_M}. Since, in the above equation only the matter degrees of freedom are involved,  one can replace $|\mathcal{M}_{\mathcal{G}_0}\rangle$ by $|\mathcal{M}_\mathcal{G}\rangle$ to reach Eq.~\ref{eq:OPDM_eff}. In this way, the long time dynamics of the occupation spectrum, evolving under $H(\lambda,V,0)$, can be recast just through simulating the matter Hamiltonian dynamics $H^{\{\lambda_i\}}_{\mathcal{M}}$ after performing disorder sampling and then diagonalizing the corresponding time averaged OPDM. 

It is also instructive to show that the diagonal ensemble of the matter Hamiltonian~(\ref{eq:disordered_H}) can also describe the asymptotic behavior of the OPDM. To this end, we expand the initial state in terms of the many body eigenstates of the original Hamiltonian~(\ref{eq:H_M}):
\begin{align}
|\Psi_0\rangle =  \sum_{n}{a_n |\Psi_n\rangle}=\sum_{k,l}{a_{kl} |\mathcal{G}_k\rangle \otimes |\mathcal{M}_l\rangle},
\end{align}
where $|\mathcal{G}_k\rangle$s represent different configurations of the gauge fields $\{g_i\}^{(k)}$, and $|\mathcal{M}_l\rangle$s are the eigenstates of the corresponding disordered matter Hamiltonian. By doing so, the time averaged OPDM takes the form, 
\begin{align} 
\bar{\rho}_{ij} = \sum_{k,l,m,\nu} & {\overline{e^{-i(E_{kl}-E_{m \nu})t}} a_{kl}^* a_{m \nu}} \times \nonumber \\ 
&\langle\mathcal{G}_k| \otimes \langle\mathcal{M}_l|c_i^{\dagger}c_j|\mathcal{G}_{m}\rangle \otimes |\mathcal{M}_{\nu}\rangle.
{\label{eq:DOPDM}}
\end{align}
Since different gauge sectors are orthogonal to each other in Eq.~\ref{eq:DOPDM}, the terms with $k=m$ have only a nonzero contribution. Moreover, for $l \neq \nu$ the oscillation of the phase $e^{-i(E_{kl}-E_{k \nu})t}$ leads to a zero time average. 
In view of the fact that the $c$-fermions are gauge invariant  and the sum over $k$ can be replaced by $\{g_i\}^{(k)}$ (equivalent to averaging over different configurations of the  gauge fields), one can obtain the time averaged OPDM by only considering the diagonal ensemble of the matter Hamiltonian~(\ref{eq:H_M}),
\begin{align}
\bar{\rho}_{ij} = \sum_{\{g_i\}} \sum_{l} { |a_{l}^{\{g_i\}}|^2  \langle\mathcal{M}_l^{\{g_i\}}|c_i^{\dagger}c_j |\mathcal{M}_{l}^{\{g_i\}}\rangle,}
\end{align}
where $|\mathcal{M}_{l}^{\{g_i\}}\rangle $s are eigenstates of  $H^{\{\lambda_i\}}_{\mathcal{M}}$ corresponding to disorder realization $\{g_i\}$, and $|a_{l}^{\{g_i\}}|^2$s are their overlap with the initial state.

\begin{figure}[t!]
\centering
\includegraphics[width=0.8\linewidth]{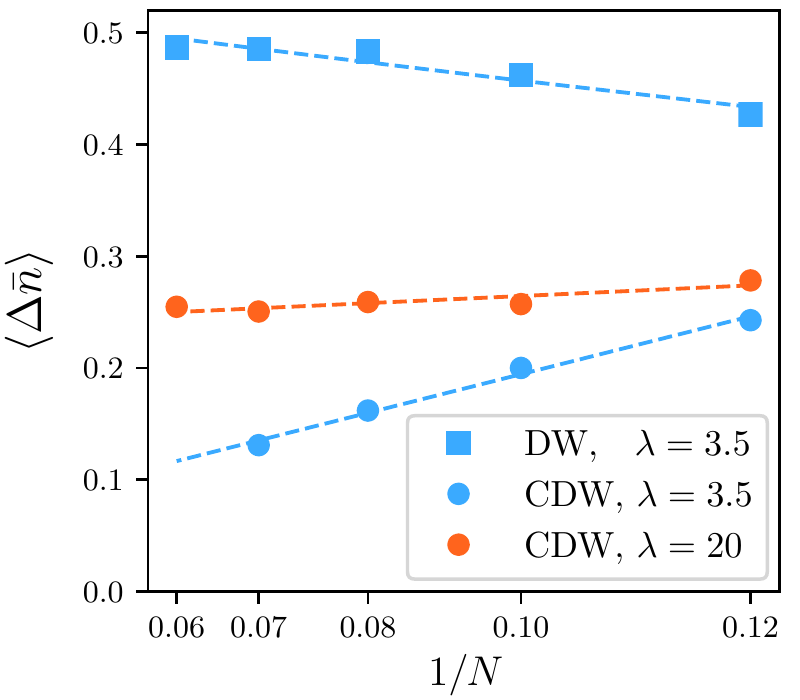}
\caption{Comparison between OPDM discontinuity of the DW and CDW prequench fermionic pattern in moderate ($\lambda=3.5$) and strong  ($\lambda=20$) interaction regimes.}
\label{fig:APP} 
\end{figure}
\section{Discontinuous OPDM spectrum deep in the SBL phase}\label{App:deep_SBL_OPDM}
In this section, we investigate the behavior of the steady state OPDM in the deep SBL phase of both DW and CDW initial states. As shown in Fig.~\ref{fig:APP}, for strong interactions, e.g., $\lambda=20$, the average discontinuity $\langle \Delta \bar{n} \rangle$ starting from the CDW state does not vanish in the thermodynamics limit, which is similar to that of the DW pre-quench state at moderate interactions even though with smaller magnitude. 
Moreover, this result is in accordance with the bubble neck structure of the SBL spectrum. Since in the infinite randomness limit, the probability of SBL eigenstates having long thermal bubbles should be statistically suppressed with system size~\cite{Li:2017}, the non-ergodic nature of the dynamics for the DW prequench state (with the largest insulating block) in moderate interactions becomes more pronounced than the one corresponding to CDW pattern even at strong interacting regime. Thus, the size of thermal bubbles within the initial fermionic configuration can serve as an effective temperature. 

Finally, it should be noted that we do not expect to observe the exponentially vanishing gap along with the smearing effects suggested in Ref.~\onlinecite{Lezama:2017}. There, these behaviors
are justified by considering the spectrum of the OPDM in the level of each individual random realization. 
In contrast, here we deal with dynamics of a disorder-free Hamiltonian and the effective OPDM spectrum extracted from an individual disorder realization has no physical meaning (see Appendix.~\ref{App:matter_OPDM}). 
%

\begin{thebibliography}{108}
\bibitem [{not()}]{note:1}%
  \BibitemOpen
  \href@noop {} {}\bibinfo {note} {We stress that, here, taking the eigenstates
  of the Hamiltonian dynamics Eq.~\ref{eq:disordered_H} in order to settle
  pertinent reference eigenstates, is irrelevant.}\BibitemShut {Stop}%
\bibitem [{foo()}]{footnote}%
  \BibitemOpen
  \href@noop {} {}\bibinfo {note} {This should be noted that, starting from
  different configurations of random $Z$-polarized initial states leads to the
  qualitatively similar dynamics. Hence, in order to reduce the computational
  cost, we focus on the dynamics obtained from just one
  configuration.}\BibitemShut {Stop}%
\end{thebibliography}

\end{document}